\documentclass[aps,groupedaddress,amssymb,showpacs,amsmath,superscriptaddres,amsbsy,amsfonts,pra]{revtex4-2}
\usepackage{amssymb,amsmath,amsthm,color,graphicx,times,graphicx}

\usepackage{subfigure}
\usepackage{times}
\usepackage{bbold}
\usepackage{color}
\usepackage[rightcaption]{sidecap}
\usepackage{graphicx}

\usepackage{tikz}
\usetikzlibrary{arrows}
\usepackage{capt-of}

\usepackage{hyperref}
\usepackage{subfigure}
\usepackage{times}
\usepackage{bbold}
\usepackage{color}
\usepackage{array}
\usepackage[french]{babel}
\usepackage{array,multirow,makecell}

\theoremstyle{plain}

\theoremstyle{definition}

\begin{document}
\title{Analytical techniques in single and multi-parameter quantum estimation theory: a focused review}
\author{Abdallah Slaoui}\email{abdallah.slaoui@um5s.net.ma}\affiliation{LPHE-Modeling and Simulation, Faculty of Sciences, Mohammed V University in Rabat, Rabat, Morocco,}\affiliation{Centre of Physics and Mathematics, CPM, Faculty of Sciences, Mohammed V University in Rabat, Rabat, Morocco.}
\author{Lalla Btissam Drissi}\email{b.drissi@um5r.ac.ma}\affiliation{LPHE-Modeling and Simulation, Faculty of Sciences, Mohammed V University in Rabat, Rabat, Morocco,}\affiliation{Centre of Physics and Mathematics, CPM, Faculty of Sciences, Mohammed V University in Rabat, Rabat, Morocco.}
\author{El Hassan Saidi}\email{e.saidi@um5r.ac.ma}\affiliation{LPHE-Modeling and Simulation, Faculty of Sciences, Mohammed V University in Rabat, Rabat, Morocco,}\affiliation{Centre of Physics and Mathematics, CPM, Faculty of Sciences, Mohammed V University in Rabat, Rabat, Morocco.}
\author{Rachid Ahl Laamara}\email{r.ahllaamara@um5r.ac.ma}\affiliation{LPHE-Modeling and Simulation, Faculty of Sciences, Mohammed V University in Rabat, Rabat, Morocco,}\affiliation{Centre of Physics and Mathematics, CPM, Faculty of Sciences, Mohammed V University in Rabat, Rabat, Morocco.}

\begin{abstract}
	
As we enter the era of quantum technologies, quantum estimation theory provides an operationally motivating framework for determining high precision devices in modern technological applications. The aim of any estimation process is to extract information from an unknown parameter embedded in a physical system such as the estimation converges to the true value of the parameter. According to the Cramér-Rao inequality in mathematical statistics, the Fisher information in the case of single-parameter estimation procedures, and the Fisher information matrix in the case of multi-parameter estimation, are the key quantities representing the ultimate precision of the parameters specifying a given statistical model. In quantum estimation strategies, it is usually difficult to derive the analytical expressions of such quantities in a given quantum state. This review provides comprehensive techniques on the analytical calculation of the quantum Fisher information as well as the quantum Fisher information matrix in various scenarios and via several methods. Furthermore, it provides a mathematical transition from classical to quantum estimation theory applied to many freedom quantum systems. To clarify these results, we examine these developments using some examples. Other challenges, including their links to quantum correlations and saturating the quantum Cramér-Rao bound, are also addressed.

\vspace{0.25cm}
\end{abstract}
\date{\today}

\maketitle

\section{Introduction}
During the last two decades, quantum information has proven, both theoretically \cite{Wilde2013} and experimentally \cite{Nagali2009}, to have made several advances and to provide the key tools for a modern technological revolution. These advances are mainly due to the prospects of quantum communications \cite{Monroe2002,Gisin2007,Zoller2005} and quantum computers \cite{Duan2000,Nielsen2002,Ladd2010,Knill2005} that allow to storage, process and transform information using quantum mechanical systems. Besides, several quantities of interest in quantum information theory are not completely accessible via observation and cannot be directly evaluated by quantum measurements \cite{Caves1981,Fujiwara2008,Giovannetti2011}. One may quote for instance; a quantum phase \cite{Ahn2000}, the purity of a quantum state \cite{Devetak2005} or quantum correlations \cite{Li2007}. In all these cases, one must resort to indirect measurement and deduce the value of the quantity of interest from its inuence on a given source. The canonical way to deal with this problem is to use the tools of quantum estimation theory \cite{Giovannetti2004,Anisimov2010}. For example, during the famouse LIGO experiment \cite{Abbott2009}, the gravitational wave inserted a relative phase into the light source which is not a directly measurable quantity, however this phase was estimated through the interference pattern observed.\par

Quantum estimation theory, as the mathematical language of quantum metrology \cite{Paris2009,Escher2011,Demkowicz-Dobrzanski2012}, is of critical importance in the development of high-precision devices in several technological areas. Its main objective is to realize high precision measurements by estimating the unknown parameters that specify a given quantum system using quantum effects. In this sense, it aims at developing new methods to improve the accuracy limit of physical parameters beyond classical metrological methods \cite{Alipour2014,Knysh2011,Tsang2013,Demkowicz-Dobrzanski2014}. In short, it is interested in achieving the highest possible accuracy in various parameter estimation tasks, and in finding measurement schemes that achieve this accuracy. Originally, metrology focused on measurements made using classical systems, such as mechanical systems described by classical physics or optical systems modeled by classical wave optics \cite{Watanabe2014}. We must therefore ask how can we deduce the limits of accuracy of parameter estimation, as well as methods that can improve accuracy in quantum systems? Is there a fundamental limit? In the classical scheme, the first answers appeared around the 1940s with the seminal paper of Rao \cite{Rao1945} and Cramer \cite{Cramer1946}, who independently found a lower bound on the variance of an arbitrary estimator. These results were extended to the multiparametric strategy by Darmois \cite{Darmois1945}. This limit, generally called the Cramér-Rao bound, is closely linked to the Fisher information, introduced by Fisher in the 1920s \cite{Fisher1923}. The Fisher information thus plays a central role in estimation theory. Its maximization on all possible quantum measures defines the quantum Fisher information and provides a quantum lower bound to the Cramér-Rao bound \cite{Braunstein1994,Braunstein1996}.\par

The quantum estimation theory is the mathematical framework for approaching the optimization problem of a quantum measurement. It applies to situations where we need to predict the value of a parameter by performing a repeated series of measurements on identical preparations of the system and then processing the data to estimate the value of the unknown parameter \cite{Paris2009}. To achieve all these tasks, we would need quantum phenomena as resources, such as quantum entanglement \cite{Einstein1935,Horodecki2009}, quantum discord \cite{Ollivier2001,Shaukat2020} and quantum coherence \cite{Streltsov2017,Slaoui2020}, to improve the sensitivity of a system. At the same time, quantum systems place intrinsic restrictions on the sensitivity that can be achieved, for example through Heisenberg uncertainty relations, which state that complementary variables cannot be measured simultaneously with unlimited precision. To determine how quantum properties enhance but also restrict the sensitivity of the system, we model its dynamics and focus on a key quantity which is the quantum Fisher information.\par

According to the quantum Cramér-Rao bound, a higher accuracy is obtained for small variances, which correspond to the largest values of the quantum Fisher information. Thus, the primary aim of all quantum metrology protocols is to reach the smallest value of the variance. However, recent studies have revealed broad connections between quantum Fisher information and other aspects of quantum mechanics, including characterization of quantum correlations \cite{Kim2018}, quantum phase transition \cite{Ye2016}, entanglement control \cite{Chapeau-Blondeau2017}, quantum speed limit \cite{Taddei2013} and quantum thermodynamics \cite{Hasegawa2020}. These links indicate that it is more than a concept in quantum metrology, but rather a fundamental quantity in quantum mechanics. Besides, it is well known that the achievable phase estimation accuracy, in the classical optical interferometry theory, is bounded by the shot noise. Then the phase uncertainty can only reach the accuracy scale $1/\sqrt{N}$, where $N$ stands for the average photon number of the light field. This phase uncertainty limit is called the standard quantum limit \cite{Giovannetti2004}. Utilizing the non-classical states of light, which in principle contains a non-zero amount of correlations, the quantum metrology theory allows to beat the shot noise and the accuracy of the phase shift estimation in an optical interferometer. The use of NOON states for example, which are maximally entangled states, can lead to a scaling of the precision $1/N$. This bound being called the Heisenberg limit \cite{Pezze2009}. Several reasons explain why quantum systems are able to outperform classical devices, and perhaps most importantly, the classical measurement boundary can be exceeded using quantum systems.\par

Quantum entanglement underlies many fundamental quantum tasks \cite{Plenio2014,Guhne2009} and is often considered synonymous with quantum correlations in early studies, although it is now recognized that the notion of quantum correlations is much broader in scope, and that entanglement is a particular, though most important, type of quantum correlation, i.e., quantum entanglement can be identified as nonlocal quantum correlations \cite{Werner1989}. The study of entanglement explicitly dates back to the seminal work of Einstein, Podolsky and Rosen \cite{Einstein1935}, and of Schrödinger \cite{Schrodinger1935,Schrodinger1936} as early as the 1930s. Today, entanglement is considered a key resource of quantum information and is often linked to quantum non-locality \cite{Bell1987,Brunner2014}. In 2002, after studying the correlation between the device and the system during a measurement, Ollivier and Zurek \cite{Ollivier2001} realized that separable states, as defined by Werner \cite{Werner1989}, can still have some correlation in the sense that they can be perturbed by local measurements. This novel measure of quantum correlations beyond entanglement is called quantum discord \cite{Luo2008,Luo2010,SlaouiShaukat2018}. They found that entanglement is not the only quantum correlation that has no classical counterpart. Other types of quantum correlations, such as quantum discord, may also be responsible for speeding up some quantum algorithms while entanglement may disappear or be negligible \cite{Modi2012}. Since the first statements of this concept, many research efforts have been devoted to understanding the mathematical properties and physical meanings of discord and similar quantities. Comprehensive review of the properties of quantum discord is available in \cite{Adesso2016} and the Refs.\cite{Brodutch2017,Adesso12016} provide recent insights into this field. Indeed, the investigation of quantum discord presents many challenges and open questions. A major difficulty with discord-like quantities is that they are difficult to calculate or to analyze. Formally, quantum discord is defined as the difference between the quantum mutual information (quantifying the total correlation) and the maximum amount of mutual information accessible to a quantum measurement (quantifying the classical correlation). To quantify the locally accessible (classical) correlations, this quantification involves an optimization over all possible local measurements. This measure is chosen to maximize the classical correlations. Due to the complexity of the optimization process, the computation of the quantum discord and its variants (such as the geometric discord) is not an easy task and the analytical results are only known for some restricted families of states. To overcome this problem, several other versions of discord have also been proposed, including linear quantum discord \cite{Ma2015}, local quantum uncertainty \cite{Girolami2013,SlaouiD2019}, and local quantum Fisher information \cite{Kim2018}. Most of these quantum correlation quantifiers for pure quantum states can be coincidental and can sometimes exhibit similar behavior for mixed states.\par

Besides, quantum metrology is a mechanism that uses the characteristic resources of quantum mechanics, such as entanglement and quantum discord, to improve the parameter estimation precision via quantum measurements beyond its classical counterpart limit. Therefore, it can be argued that quantum resources are very important in all quantum metrology protocols \cite{Giovannetti2011}. In this review, our concerns include the intrinsic connections between quantum correlations and estimation theory and their roles in improving estimated parameter precision. For this purpose, this review is presented with many technical details, which could help researchers to follow and better understand the corresponding results. Further to presenting an overview of the main developments in estimation theory, we try to summarize and reformulate some of the calculations scattered in a large number of works, combined of course with our own results. To begin with, we introduce the basic concepts of classical estimation theory before discussing the quantum advantage. We then derive the classical Fisher information as a measure of information that can be related to the variance of an estimated parameter. We then show how classical Fisher information can be generalized for all possible measures to quantum Fisher information. The motivation for studying these quantities becomes clear when we introduce and derive the Cramér-Rao bound, which relates the quantum and classical Fisher information to the variance. In fact, this Cramér-Rao quantum bound is always reached in saturation in the case where only one parameter is estimated, and in this situation, the quantum correlations help us to improve the accuracy and efficiency of quantum metrology protocols. Conversely, this limit is hard to saturate in the case of multiparametric estimation due to the incompatibility between the optimal measures of the different estimated parameters. The pertinent object in the multiparametric estimation problem is given by the so-called quantum Fisher information matrix. To this end, in the second part of this review, we provide the complete techniques on the computation of this matrix, and we show that the simultaneous multi-parameter strategy is always advantageous and can provide better precision than the individual strategy in multiparameter estimation procedures. Further, we intend to focus our attention on the interaction between quantum correlations and quantum estimation theory and the characterization of quantum correlations in terms of quantum Fisher information will be discussed.

\section{Basic concepts of classical estimation theory}
Estimation theory is an influential mathematical concept applied in various communication and signal processing applications. This theory is useful for estimating the desired information in received data and is therefore used in a variety of applications ranging from radar to speech processing. In this section, we will list the basic statistical concepts as well as the estimation process. As we have mentioned, the main mathematical tools used by metrology belong to statistics. In addition, we are particularly interested in estimation theory, which shows how to correctly estimate a quantity from a data sample \cite{Kay1993}. The data can be of any type. For example, the data sample can be a set of the results of random draws, or even the wavelengths of photons coming out of a radioactive sample. Assume that a non-deterministic process produces a set of data $x=\left\lbrace x_{1},x_{2},...,x_{n}\right\rbrace$, which is a particular occurrence (or realization) of a set of independent and identically distributed random variables denoted by $X=\left\lbrace X_{1},X_{2},...,X_{n}\right\rbrace$. The probability density function corresponding to any random variable in this set is $p\left(x;\theta\right)$, where $x\in R$ is a possible realization of a random variable and  $\theta$ is a real number (e.g., it could be a physical parameter, such as the phase in an optical interferometer) that is parametrized by $p\left(x;\theta\right)$. Hence, the probability that a measurement result in our data set lies in the interval $x_{0}\leq x\leq x_{f}$ is determined by the integral of $p\left(x;\theta\right)$ as: $\int_{x_{0}}^{x_{f}}p\left(x;\theta\right)dx$.\par

Generally, the central problem in parameter estimation is the following \cite{Kay1993,Volz2008}: For a data set $x$ of $n$ test results, how accurately can we predict the value of $\theta$? Indeed, in physics, the parameter $\theta$ is a physical property of the system studied and we want to determine its value from the experimental data $x_{i}$. For this purpose, an estimator is used to transform the data $x$ into a prediction of $\theta$. We will denote such an estimator by $\boldsymbol{\hat{\theta}}\left(x \right)$, which happens to be a random variable since it depends on the values taken from the random variables in $X$. In other words, estimating the parameter $\theta$ consists in giving an approximate value to this parameter from a population survey. This implies that the true value of $\theta$ is obtained if $\theta$ coincides with the mathematical expectation of its estimator $\boldsymbol{\hat{\theta}}\left(X\right)$, i.e. $\theta=\textbf{E}\left[\boldsymbol{\hat{\theta}}\left(X \right)\right]$.

\begin{SCfigure}[0.9][h]
\includegraphics[scale=0.05]{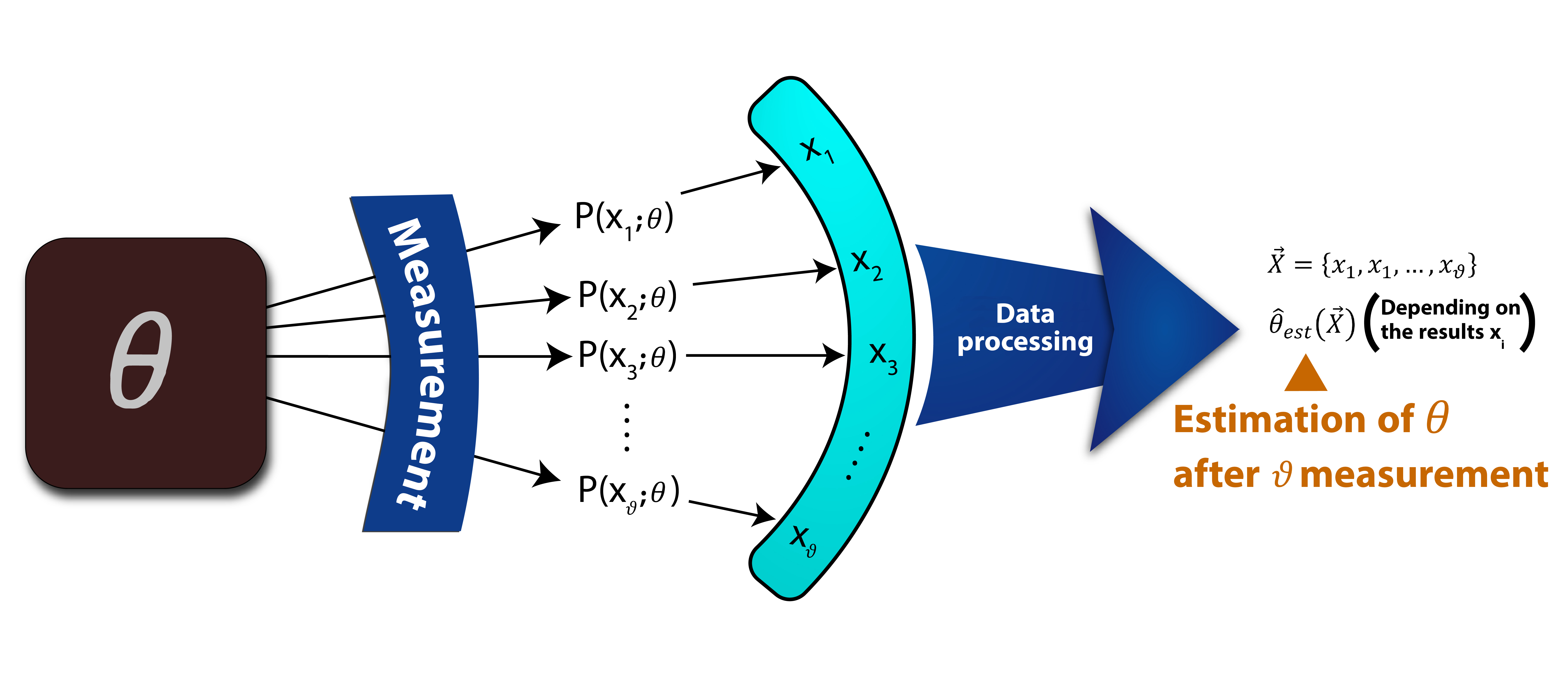}
\caption{A schematic illustration of a complete classical metrological process}
\end{SCfigure}
We can now reformulate the estimation problem as follows; For any set of data $x$, what is the small value of the error of our estimator? More precisely, the minimum error is given by the variance of this estimator ${\rm Var}_{\theta}\left(\hat{\theta}\right)$ \cite{Watanabe2014}. It is defined by
\begin{align}
	{\rm Var}_{\theta}\left(\hat{\theta}\right)&:=\textbf{E}\left[\left(\boldsymbol{\hat{\theta}}\left(x\right) -\theta\right)^{2}\right]\notag\\&=\int p\left(x_{1};\theta\right)p\left(x_{2};\theta\right)...p\left(x_{n};\theta\right)\left(\boldsymbol{\hat{\theta}}\left(x\right)-\theta\right)^{2}dx_{1}dx_{2}...dx_{n}\notag\\&=\int p\left(x;\theta\right)\left(\boldsymbol{\hat{\theta}}\left(x\right)-\theta\right)^{2}dx^{n},\label{Delta}
\end{align}
which is used to quantify the performance of our estimation task, with $\textbf{E}$ being a mathematical expectation. In the above equation (\ref{Delta}), we employed the fact that $dx^{n}=dx_{1}dx_{2}...dx_{n}$ and $p\left(x;\theta\right)=p\left(x_{1};\theta\right)p\left(x_{2};\theta\right)...p\left(x_{n};\theta\right)$.\par 
It is therefore clear that the law of $X$ will be completely known if the probability distribution function $p\left(x;\theta\right)$ is known, this necessarily implies the knowledge of the value of the unknown parameters $\theta$. In this context, the estimation theory aims at approximating numerically the value of the unknown parameters without knowing the probability density and the probability law which governs our problem. In this formalism, we make the implicit assumption that $\theta$ is knowable with $\boldsymbol{\hat{\theta}}\left(x\right)$ and have access to a set of data $x$.
\subsection{Classical statistical schemes}
Consider a random experiment whose results are described by a random variable $X$, with a probability space $\left(X,x,p\left(x\right) \right)$ and a probability density $p\left(x\right)$. The task here is to reconstruct $p\left(x\right)$, which is called the genuine probability density, from $N$ independent sample positions or observations of $X$. There are many ways to approach the learning problem of $p\left(x\right)$, but if the functional form of $p\left(x\right)$ is already known, or can be assumed with reasonable accuracy, a parametric approach is quite natural. The real probability density $p\left(x\right)$ is assumed to belong to a parametric family of probability densities $\left\lbrace p\left(x;\theta\right)\right\rbrace_{\theta\in\Theta}$ where $\Theta\subset\mathbb{R}^{n}$ is the parameter space. A classical statistical model $S_{C}$ is a family of probability densities on $X$ parametrized by $n$ real parameters $\theta\in\Theta\subset\mathbb{R}^{n}$ and noted by
\begin{equation}
	S_{C}=\left\lbrace \forall x\in X; p\left(x;\theta\right),\hspace{0.25cm}\left(X,x,p\left(x\right) \right)\mapsto\Theta,\hspace{0.25cm}\theta\equiv\left(\theta^{1},\theta^{2},...,\theta^{n}\right)\in\Theta\right\rbrace,\label{MS}
\end{equation}
where the parametrization application $\left(X,x,p\left(x\right) \right)\mapsto\Theta$ is injective and all possible derivatives $\partial_{\theta^{1}}...\partial_{\theta^{n}}p\left(x;\theta\right)$ exist, i.e. we can be derivable $n^{th}$ times with respect to the estimated parameters. It is interesting to note that the probability density $p\left(x\right)$ is always positive and normalized. So we have two cases; If $X$ is countable, then $p\left(x;\theta\right)$ is described in a discrete and normalized space such that  
\begin{equation}
	\sum_{x\in X}p\left(x;\theta\right)=1, \hspace{1cm}\forall \theta\in\Theta.
\end{equation}
If $X$ is uncountable, then $p\left(x;\theta\right)$ is described in a continuous and normalized space like
\begin{equation}
	\int_{x}p\left(x;\theta\right)dx=1, \hspace{1cm}\forall \theta\in\Theta.
\end{equation}
To simplify our discussion, we employ the latter notation throughout this review.
\subsection{Score, Fisher Information and Cramér-Rao theorem}
Measurements on physical systems produce probabilistic results, and the estimation of physical parameters, describing a system or governing its evolution, is a problem involving statistical inference. The source of statistical errors may be related to imperfections in the experiment or in the measurements, or it may be more fundamental, imposed, for example, by the Heisenberg uncertainty relations in a quantum configuration. In the classical case, a quantity that is key to this discussion is the Score which we denote by ${\cal L}_{\theta}\left(x\right)$. The Score is the partial derivative of the natural logarithm of the likelihood function, and thus indicates the sensitivity of the system to an infinitesimal variation in the value of the parameter $\theta$. It is defined as \cite{Cox1979}
\begin{equation}
	{\cal L}_{\theta}\left(x\right) =\frac{\partial}{\partial \theta}\log p\left(x;\theta\right),
\end{equation}
which quantifies the sensitivity of the probability density function $p\left(x;\theta\right)$ to the variations of the parameter $\theta$. This arises from the fact that
\begin{align}
	\textbf{E}\left({\cal L}_{\theta}\left(x\right)\right)&=\int p\left(x;\theta \right)\frac{\partial}{\partial \theta}\log p\left(x;\theta\right)dx\notag\\&=\int p\left(x;\theta \right)\frac{1}{p\left(x;\theta \right)}\frac{\partial}{\partial \theta}p\left(x;\theta \right)dx\notag\\&=\int \frac{\partial}{\partial \theta}p\left(x;\theta \right)dx=0.\label{46}
\end{align}
The classical Fisher information ${\cal F}\left(\theta\right)$ in a single observation of a probability density function $p\left(x;\theta\right)$ is the variance of the Score ${\cal L}_{\theta}\left(x\right)$, namely
\begin{align}
	{\cal F}\left(\theta\right)&:= {\cal F}\left( p\left(x;\theta\right)\right):={\rm Var}\left( {\cal L}_{\theta}\left(x\right)\right)\notag\\&=\textbf{E}\left[\left(\frac{\partial}{\partial \theta}\log p\left(x;\theta\right) \right)^{2}\right]\notag\\&=\int p\left(x;\theta \right)\left(\frac{\partial}{\partial \theta}\log p\left(x;\theta\right) \right)^{2}dx.\label{FC}
\end{align}
Formally, the Fisher information (\ref{FC}) quantifies the information provided by a sample on the parameter $\theta$. In addition, a Fisher information close to zero indicates a sample with little information on the value of $\theta$. We note here an important and useful property of classical Fisher information; namely, non-negativity and additivity for uncorrelated processes. In other words, if $p\left(\boldsymbol{x};\theta\right)=p\left(x_{1};\theta\right)p\left(x_{2};\theta\right)...p\left(x_{n};\theta\right)$, then
\begin{align}
	{\cal F}\left(p\left(\boldsymbol{x};\theta\right)\right)&={\rm Var}\left[\frac{\partial}{\partial \theta}\log p\left(x_{1};\theta\right)+...+\frac{\partial}{\partial \theta}\log p\left(x_{n};\theta\right)\right]\notag\\&=n{\rm Var}\left[\frac{\partial}{\partial \theta}\log p\left(x_{1};\theta\right) \right]=n{\cal F}\left(\theta\right).
\end{align}
Therefore, the Fisher information in a random sample is just $n$ times the information in a single observation. This should make intuitive sense because it means that more observations produce more information.\par

The Cramér-Rao bound provides a limit on how the estimation can be performed. It sets a fundamental limit on the variance of the estimator $\boldsymbol{\hat{\theta}}$ regardless of how it is constructed. It is expressed by the following formula \cite{Rao1945,Cramer1946}
\begin{equation}
	{\rm Var}\left(\boldsymbol{\hat{\theta}}\right)\geq \frac{1}{n{\cal F}\left(\theta\right)},\label{BCR}
\end{equation}
where $n$ is the trial number (the number of entries in the data set $\boldsymbol{x}$). This bound (\ref{BCR}) asserts that the error of any unbiased estimator $\boldsymbol{\hat{\theta}}$ of $\theta$ is bounded by the inverse of the Fisher information. This indicates that the variance of an estimator $\boldsymbol{\hat{\theta}}$ is at least as high as the inverse of the Fisher information, i.e., the greater the Fisher information, the higher the estimation accuracy can be expected. Thus, Fisher information is the quantity that produces an ultimate limit on the precision that can be achieved in a parameter estimation problem. In this way, the Fisher information captures the maximal information that can be extracted about the actual value sought for the parameter $\theta$. Moreover, one can obviously improve the precision by repeating the measurement (as the expression of the Cramér-Rao bound reflects with the factor $1/n$) provided that one ensures the independence of the repetitions. As we have seen, this bound is asymptotically attainable by choosing the maximum estimator. For example, if the estimator is the mean value, any Gaussian probability distribution $p\left(\boldsymbol{X};\theta\right)$ saturates the inequality. An estimator for which equality is satisfied is said to be efficient, but with a finite number of measurements, there is no guarantee that such an estimator exists.\par

We now provide a proof of the Cramér-Rao limit following an approach similar to the Ref.\cite{Kok2010}. In order to prove this (\ref{BCR}), we have to make the assumption that the estimator $\boldsymbol{\hat{\theta}}\left(x\right)$ must be unbiased. This means that on average, the estimator $\boldsymbol{\hat{\theta}}$ is equal to the true value of $\theta$;
\begin{equation}
	\textbf{E}\left(\hat{\theta}\left(\boldsymbol{x} \right)\right)\equiv\theta.\label{biaise}
\end{equation}
By shifting $\theta$ to the left side and writing the average explicitly in terms of the probability density function $p\left(\boldsymbol{x};\theta\right)$. The above equation (\ref{biaise}) becomes
\begin{align}
	\int p\left(\boldsymbol{x};\theta\right)\left(\textbf{E}\left(\hat{\theta}\left(\boldsymbol{x} \right)\right)-\theta \right)d\boldsymbol{x}^{n}=0. 
\end{align}
If we derive this expression with respect to $\theta$, we obtain
\begin{align}
	\frac{\partial}{\partial\theta}\int p\left(\boldsymbol{x};\theta\right)\left(\textbf{E}\left(\hat{\theta}\left(\boldsymbol{x} \right)\right)-\theta \right)d\boldsymbol{x}^{n}=\int \frac{\partial}{\partial\theta}\left(p\left(\boldsymbol{x};\theta\right) \right)\left(\textbf{E}\left(\hat{\theta}\left(\boldsymbol{x} \right)\right)-\theta \right)d\boldsymbol{x}^{n}-\int p\left(\boldsymbol{x};\theta\right)d\boldsymbol{x}^{n}=0,\label{412}
\end{align}
where we used the fact that the estimator doesn't depend explicitly on $\theta$, hence $\frac{\partial}{\partial\theta}\left(\hat{\theta}\left(\boldsymbol{x} \right)-\theta \right)=-1$. Also, for any probability density function we have $\int p\left(\boldsymbol{x};\theta\right)d\boldsymbol{x}^{n}=1$. Therefore, the equation (\ref{412}) yields
\begin{align}
	\int \frac{\partial p\left(\boldsymbol{x};\theta\right)}{\partial\theta}\left(\hat{\theta}\left(\boldsymbol{x} \right)-\theta \right)d\boldsymbol{x}^{n}=1.\label{413}
\end{align}
It should be noted that $\partial p\left(\boldsymbol{x};\theta\right)/\partial\theta$ can still be written as follows
\begin{align}
	\frac{\partial}{\partial\theta}\left(\prod_{k=1}^{n}p\left(x_{k};\theta\right) \right)&=\frac{\partial p\left(x_{1};\theta\right)}{\partial\theta}\prod_{k\neq1}^{n}p\left(x_{k};\theta\right)+\frac{\partial p\left(x_{2};\theta\right)}{\partial\theta}\prod_{k\neq2}^{n}p\left(x_{k};\theta\right)+...+\frac{\partial p\left(x_{n};\theta\right)}{\partial\theta}\prod_{k\neq n}^{n}p\left(x_{k};\theta\right)\notag\\&=\sum_{l=1}^{n}\left[ \frac{\partial p\left(x_{l};\theta\right)}{\partial\theta}\prod_{k\neq l}^{n}p\left(x_{k};\theta\right)\right],
\end{align}
where $\prod_{k\neq l}^{n}p\left(x_{k};\theta\right)$ denotes the multiplication of all terms $p\left(x_{k};\theta\right)$ for $k\in\left\lbrace 1,2,...,n\right\rbrace$ except for the $l$-th term. When writing $\prod_{k\neq l}^{n}p\left(x_{k};\theta\right)=p\left(\boldsymbol{x};\theta\right)/p\left(x_{l};\theta\right)$, we get
\begin{align}
	\frac{\partial p\left(\boldsymbol{x};\theta\right)}{\partial\theta}=\sum_{l=1}^{n}\frac{1}{p\left(x_{l};\theta\right)}\frac{\partial p\left(x_{l};\theta\right)}{\partial\theta}p\left(\boldsymbol{x};\theta\right)=p\left(\boldsymbol{x};\theta\right)\sum_{l=1}^{n}\frac{\partial\log p\left(x_{l};\theta\right)}{\partial\theta}.
\end{align}
The previous equation could be used to re-express equation (\ref{413}) as follows
\begin{align}
	\int p\left(\boldsymbol{x};\theta\right)\sum_{l=1}^{n}\frac{\partial\log p\left(x_{l};\theta\right)}{\partial\theta}\left(\hat{\theta}\left(\boldsymbol{x} \right)-\theta \right)d\boldsymbol{x}^{n}&=\int\left[\sqrt{p\left(\boldsymbol{x};\theta\right)}\sum_{l=1}^{n}\frac{\partial\log p\left(x_{l};\theta\right)}{\partial\theta} \right]\notag\\&\times\left[\sqrt{p\left(\boldsymbol{x};\theta\right)}\left(\hat{\theta}\left(\boldsymbol{x} \right)-\theta \right) \right]d\boldsymbol{x}^{n}=1, \label{416}
\end{align}
where we decided to write the probability density function $p\left(\boldsymbol{x};\theta\right)$ as a product of its square roots. Thus the equation (\ref{416}) takes the following form
\begin{equation}
	\int f\left(\boldsymbol{x}\right) g\left(\boldsymbol{x} \right)d\boldsymbol{x}^{n}=1,
\end{equation}
with
\begin{equation}
	f\left(\boldsymbol{x} \right)\equiv\sqrt{p\left(\boldsymbol{x};\theta\right)}\sum_{l=1}^{n}\frac{\partial\log p\left(x_{l};\theta\right)}{\partial\theta},\hspace{1cm}{\rm et}\hspace{1cm}g\left(\boldsymbol{x} \right)\equiv\sqrt{p\left(\boldsymbol{x};\theta\right)}\left(\hat{\theta}\left(\boldsymbol{x} \right)-\theta \right). 
\end{equation}
We are going to use now the Cauchy-Schwarz inequality \cite{Helstrom1976}:
\begin{equation}
	\left(\int f\left(\boldsymbol{x} \right) g\left(\boldsymbol{x} \right)d\boldsymbol{x}^{n}\right)^{2}\leq\int f\left(\boldsymbol{x}\right)^{2}d\boldsymbol{x}^{n}\int g\left(\boldsymbol{x}^{\prime}\right)^{2}d{\boldsymbol{x}^{\prime}}^{n}.
\end{equation}
We square the two sides of the equation (\ref{416}) and we apply the Cauchy-Schwarz inequality in order to obtain
\begin{equation}
	1\leq \left[\int p\left(\boldsymbol{x};\theta\right)\left(\sum_{l=1}^{n}\frac{\partial\log p\left(x_{l};\theta\right)}{\partial\theta}\right)^{2}d\boldsymbol{x}^{n}\right]\left[\underbrace {\int p\left(\boldsymbol{x}^{\prime};\theta\right) \left(\hat{\theta}\left(\boldsymbol{x}^{\prime} \right)-\theta \right)^{2}d{\boldsymbol{x}^{\prime}}^{n}}_{{\rm Var}\left(\boldsymbol{\hat{\theta}}\right)}\right].  
\end{equation}
The second term of the last equation is simply the variance of the estimator $\boldsymbol{\hat{\theta}}$. Then
\begin{equation}
	{\rm Var}\left(\boldsymbol{\hat{\theta}}\right)\geq\frac{1}{\int p\left(\boldsymbol{x};\theta\right)\left(\sum_{l=1}^{n}\frac{\partial\log p\left(x_{l};\theta\right)}{\partial\theta}\right)^{2}d\boldsymbol{x}^{n}}.\label{421}
\end{equation}
To complete the proof, we need to show that the denominator of the equation (\ref{421}) is equal to $n{\cal F}\left(\theta\right)$. By expanding the summation term in the denominator, we can express it as follows
\begin{align}
	\int p\left(\boldsymbol{x};\theta\right)\left(\sum_{l=1}^{n}\frac{\partial\log p\left(x_{l};\theta\right)}{\partial\theta}\right)^{2}d\boldsymbol{x}^{n}&=\int p\left(\boldsymbol{x};\theta\right)\sum_{k=1}^{n}\sum_{l=1}^{n} \frac{\partial\log p\left(x_{l};\theta\right)}{\partial\theta}\frac{\partial\log p\left(x_{k};\theta\right)}{\partial\theta}d\boldsymbol{x}^{n}\notag\\&=\int p\left(\boldsymbol{x};\theta\right)\sum_{l=1}^{n}\left( \frac{\partial\log p\left(x_{l};\theta\right)}{\partial\theta}\right)^{2}d\boldsymbol{x}^{n}\notag\\&+\int p\left(\boldsymbol{x};\theta\right)\sum_{k,l=1,k\neq l}^{n} \frac{\partial\log p\left(x_{l};\theta\right)}{\partial\theta}\frac{\partial\log p\left(x_{k};\theta\right)}{\partial\theta}d\boldsymbol{x}^{n}.\label{422}
\end{align}
The last term of this equation (\ref{422}) can be rewritten as
\begin{align}
	\int p\left(\boldsymbol{x};\theta\right)\sum_{k,l=1,k\neq l}^{n} \frac{\partial\log p\left(x_{l};\theta\right)}{\partial\theta}&\frac{\partial\log p\left(x_{k};\theta\right)}{\partial\theta}d\boldsymbol{x}^{n}=\int dx_{1}dx_{2}...dx_{n}p\left(x_{1};\theta\right)p\left(x_{2};\theta\right)...p\left(x_{n};\theta\right)\notag\\&\times\sum_{k,l=1,k\neq l}^{n} \frac{\partial\log p\left(x_{l};\theta\right)}{\partial\theta}\frac{\partial\log p\left(x_{k};\theta\right)}{\partial\theta},
\end{align}
then remove the summation and integrate some of the terms as shown below
\begin{align}
	&\sum_{k\neq1}\left(\int p\left(x_{1};\theta\right)\frac{\partial\log p\left(x_{1};\theta\right)}{\partial\theta}dx_{1}\right)\left(\int p\left(x_{k};\theta\right) \frac{\partial\log p\left(x_{k};\theta\right)}{\partial\theta}dx_{k}\right)\left(\int \prod_{m\neq1}^{n}p\left(x_{m};\theta\right)dx_{m}\right)+\notag\\& \sum_{k\neq2}\left(\int p\left(x_{2};\theta\right)\frac{\partial\log p\left(x_{2};\theta\right)}{\partial\theta}dx_{2}\right)\left(\int p\left(x_{k};\theta\right) \frac{\partial\log p\left(x_{k};\theta\right)}{\partial\theta}dx_{k}\right)\left(\int \prod_{m\neq2}^{n}p\left(x_{m};\theta\right)dx_{m}\right)+... 
\end{align}
Every term above is a product of the expectation values of the Scores. Therefore, we can rewrite this expression as
\begin{align}
	&\sum_{k\neq1}\textbf{E}\left( \frac{\partial\log p\left(x_{1};\theta\right)}{\partial\theta}\right)\textbf{E}\left( \frac{\partial\log p\left(x_{k};\theta\right)}{\partial\theta}\right)+\sum_{k\neq2}\textbf{E}\left( \frac{\partial\log p\left(x_{2};\theta\right)}{\partial\theta}\right)\textbf{E}\left( \frac{\partial\log p\left(x_{k};\theta\right)}{\partial\theta}\right)\notag\\&\hspace{2cm}+...+\sum_{k\neq n}\textbf{E}\left( \frac{\partial\log p\left(x_{n};\theta\right)}{\partial\theta}\right)\textbf{E}\left( \frac{\partial\log p\left(x_{k};\theta\right)}{\partial\theta}\right)\notag\\&= \sum_{k\neq1}\textbf{E}\left({\cal L}_{\theta}\left(x_{1}\right) \right)\textbf{E}\left({\cal L}_{\theta}\left(x_{k}\right) \right)+\sum_{k\neq2}\textbf{E}\left({\cal L}_{\theta}\left(x_{2}\right) \right)\textbf{E}\left({\cal L}_{\theta}\left(x_{k}\right) \right)+...+\sum_{k\neq n}\textbf{E}\left({\cal L}_{\theta}\left(x_{n}\right) \right)\textbf{E}\left({\cal L}_{\theta}\left(x_{k}\right) \right)\notag\\&=0.  
\end{align}
We showed earlier that the expectation value of a score is always zero (see equation (\ref{46})). Therefore, the whole expression is equal to zero and we can simplify the equation (\ref{422}) to read
\begin{align}
	\int p\left(\boldsymbol{x};\theta\right)\left(\sum_{l=1}^{n} \frac{\partial\log p\left(x_{l};\theta\right)}{\partial\theta}\right)^{2}d\boldsymbol{x}^{n}&=\int p\left(\boldsymbol{x};\theta\right)\sum_{l=1}^{n} \left(\frac{\partial\log p\left(x_{l};\theta\right)}{\partial\theta} \right)^{2}d\boldsymbol{x}^{n}\notag\\&=\sum_{l=1}^{n}\int dx_{1}...dx_{n}p\left(x_{1};\theta\right)...p\left(x_{n};\theta\right)\sum_{l=1}^{n} \left(\frac{\partial\log p\left(x_{l};\theta\right)}{\partial\theta}\right)^{2}\notag\\&=\sum_{l=1}^{n}\left[\int p\left(x_{l};\theta\right)\left(\frac{\partial\log p\left(x_{l};\theta\right)}{\partial\theta}\right)^{2}dx_{l}\right]\left[\underbrace {\int\prod_{k\neq l}^{n}p\left(x_{k};\theta\right)dx_{k}}_{=1}\right]\notag\\&=\sum_{l=1}^{n}\textbf{E}\left[\left(\frac{\partial\log p\left(x_{l};\theta\right)}{\partial\theta}\right) \right]. \label{426} 
\end{align}
Consequently, each term in equation (\ref{426}) is the classical Fisher information for the random variable $X_{l}$ with probability density function $p\left(x_{l};\theta\right)$. Since the considered random variables are independent and identically distributed, all $p\left(x_{l};\theta\right)$ are identical. As a result, we find
\begin{equation}
	\int p\left(\boldsymbol{x};\theta\right)\left(\sum_{l=1}^{n}\frac{\partial\log p\left(x_{l};\theta\right)}{\partial\theta}\right)^{2}d\boldsymbol{x}^{n}=\sum_{l=1}^{n}{\cal F}\left( p\left(x_{l};\theta\right)\right)=n{\cal F}\left(\theta\right).\label{427}
\end{equation}
Finally, by combining equation (\ref{427}) with equation (\ref{421}), we obtain the famous Cramér-Rao bound (\ref{BCR}). We will list here some properties of the classical Fisher information ${\cal F}\left(\theta\right)$: 
\begin{itemize}
	\item[($i$)]Nothing about $\theta$ can be learned when ${\cal F}\left(\theta\right)=0$, and inversely, $\theta$ can be learned perfectly when ${\cal F}\left(\theta\right)\longmapsto\infty$.
	\item[($ii$)] Classical Fisher information is a strictly positive quantity. This can be proved by examining a more general form of ${\cal F}\left(\theta\right)$ which takes into account many variables $\left\lbrace \theta_{1},...,\theta_{d}\right\rbrace$ (i.e. the case of multiparameter estimation) and by examining the form of the resulting matrices whose positive sign can be shown.
	\item[($iii$)] Unlike other information measures, such as mutual information or Shannon entropy, classical Fisher information is dimensional. We acquire units because the probability densities $p\left(x_{l};\theta\right)$ are generally dimensional and also because the derivative $\partial_{\theta}$ has units of $\theta^{-1}$. This is required to relate ${\cal F}\left(\theta\right)$ to the variance ${\rm Var}\left(\theta\right)$ through the Cramér-Rao inequality (\ref{BCR}).
	
	\item[($iv$)] Classical Fisher information is related to mutual information, but they are two fundamentally dissimilar quantities. For an intuitive understanding, ${\cal F}\left(\theta\right)$ is concerned with the probability that an estimator $\boldsymbol{\hat{\theta}}$ comes close to the "true" value of $\theta$ while the mutual information ${\cal I}\left(X:Y\right)$ measures the correlation between the random variables $X$ and $Y$.
\end{itemize}
\section{Extending to classical multiparametric metrology}
\subsection{Classical Fisher Information Matrix}
The main assignment here is to expand the estimation problem to the multiparameter case by estimating a set of unknown parameters. In this case, the various parameters we wish to estimate are constructed in the vector defined in the parameter space $\boldsymbol{ \theta}\left(\boldsymbol{x} \right)=\left(\theta_{1}\left( \boldsymbol{x}\right),...,\theta_{d}\left(\boldsymbol{x}\right)\right)^{T}\in\mathbb{R}^{d}$. Thereby, the realization of the estimation vector is represented by $\boldsymbol{\hat{\boldsymbol{ \theta}}}\left( \boldsymbol{x}\right)=\left(\hat{\theta}_{1}\left(\boldsymbol{x} \right),...,\hat{\theta}_{d}\left(\boldsymbol{x}\right)\right)^{T}$. For any given estimator $\boldsymbol{\hat{\boldsymbol{ \theta}}}\left(\boldsymbol{x}\right)$ of $\boldsymbol{\boldsymbol{\theta}}\left(\boldsymbol{x}\right)$, the mathematical expectation is given by $\textbf{E}\left[\left(\boldsymbol{\boldsymbol{\theta}}- \boldsymbol{\hat{\boldsymbol{\theta}}}\right) \left(\boldsymbol{\boldsymbol{\theta}}- \boldsymbol{\hat{\boldsymbol{\theta}}}\right)^{T}\right]$. Subsequently, the mean square error is equal to the covariance matrix ${\rm Cov}\left( \boldsymbol{\hat{\boldsymbol{ \theta}}}\right)$. One of the key results of classical probability theory, namely the Cramér-Rao inequality, enforcing a lower bound on the covariance matrix \cite{Paris2009} 
\begin{equation}
	{\rm Cov}\left( \boldsymbol{\hat{\boldsymbol{ \theta}}}\right)\geq \mathbb{F}\left(\boldsymbol{\boldsymbol{ \theta}} \right)^{-1},
\end{equation}
where $\mathbb{F}\left(\boldsymbol{\boldsymbol{ \theta}}\right)$ is the classical Fisher information matrix having the following elements
\begin{align}
	\left[\mathbb{F}\left(\boldsymbol{\boldsymbol{ \theta}} \right)\right]_{ij}=\int p\left(\boldsymbol{x};\boldsymbol{\boldsymbol{ \theta}}\right)\left(\frac{\partial\log p\left(\boldsymbol{x};\boldsymbol{\boldsymbol{ \theta}}\right)}{\partial\theta_{i}}\frac{\partial\log p\left(\boldsymbol{x};\boldsymbol{\boldsymbol{ \theta}}\right)}{\partial\theta_{j}}\right)dx,\label{CFI}
\end{align}
depending on the probability distribution $p\left(\boldsymbol{x};\boldsymbol{\boldsymbol{ \theta}}\right)$ of the results $\boldsymbol{x}$. Interestingly, the Cramér-Rao inequality only applies to probability distributions that satisfy the following regularity condition
\begin{equation}
	\textbf{E}\left[\frac{\partial\log p\left(\boldsymbol{x};\boldsymbol{\boldsymbol{ \theta}}\right)}{\partial\theta_{i}}\right]=0,\hspace{1cm}\forall \theta\in\Theta.\label{430}
\end{equation}
If such a locally unbiased estimator exists, the Cramér-Rao bound can always be saturated. However, the identification of the saturation conditions of the Cramér-Rao bound is a complex subject that involves a variety of technical details. In particular, for the probability distribution satisfying equation (\ref{430}), there exists a local unbiased estimator $\boldsymbol{\hat{\boldsymbol{ \theta}}}$ saturating the Cramér-Rao bound provided that \cite{Kay1993}
\begin{equation}
	\frac{\partial\log p\left(\boldsymbol{x};\boldsymbol{\boldsymbol{ \theta}}\right)}{\partial\theta}=\mathbb{F}\left(\boldsymbol{\boldsymbol{ \theta}} \right)\left(\boldsymbol{\hat{\boldsymbol{ \theta}}}-\boldsymbol{\boldsymbol{ \theta}}\right).
\end{equation}

\subsection{Example: Classical Estimation via Normal Distribution $X\sim N\left(\mu,\sigma\right)$}
Many practical situations can be modeled using random variables that are governed by specific laws. It is therefore important to study probabilistic models that can later allow us to analyze the fluctuations of certain phenomena by evaluating, for example, the probabilities that a particular event or outcome will be observed. In the area of statistics, normal distributions are among the most widely used probability distributions for modeling natural experiments involving multiple random occurrences \cite{Ahsanullah2014,Nadarajah2005}. They are also called Gaussian distributions \cite{Tweedie1957,Gopinath1998}. More formally, a normal distribution is an absolutely continuous probability law and depends on two unknown parameters, namely the mathematical expectation which is a real number denoted $\mu$, and its standard deviation or variance denoted $\sigma$. The probability distribution function of the normal law of expectation $\mu$ and standard deviation $\sigma$ is provided by
\begin{align}
p\left(x;\boldsymbol{ \theta}\right)=\frac{1}{\sqrt{2\pi }\sigma}\exp\left[-\frac{\left(x-\mu\right)^{2}}{2\sigma^{2}}\right]\equiv\frac{1}{\sqrt{2\pi\theta_{2}}}\exp\left[-\frac{\left(x-\theta_{1}\right)^{2}}{2\theta_{2}}\right].
\end{align}
The classical statistical model (\ref{MS}) associated with this normal distribution is given by
\begin{equation}
	S_{N\left(\mu,\sigma\right)}=\left\lbrace \forall x\in N\left(\mu,\sigma\right); p\left(x;\boldsymbol{ \theta}\right)\equiv\frac{1}{\sqrt{2\pi\theta_{2}}}\exp\left[-\frac{\left(x-\theta_{1}\right)^{2}}{2\theta_{2}}\right],\hspace{0.25cm}\Theta=\left(\mu,\sigma^{2}\right)^{T},\hspace{0.25cm}\boldsymbol{ \theta}=\left(\theta_{1}\equiv\mu,\theta_{2}\equiv\sigma^{2}\right)\in\Theta, \hspace{0.2cm}n=2\right\rbrace.
\end{equation}
It is interesting to note that
\begin{equation}
	\log p\left(x;\boldsymbol{\theta}\right)=-\frac{1}{2}\log\left(2\pi \sigma^{2}\right)-\frac{\left(x-\mu \right)^{2}}{2\sigma^{2}},
\end{equation}
we then obtain
\begin{align}
	\frac{\partial^{2}\log p\left(x;\boldsymbol{\theta}\right)}{\partial^{2}\sigma^{2}}=\frac{1}{2\sigma^{4}}-\frac{\left(x-\mu \right)^{2}}{\sigma^{6}},\hspace{2cm}\frac{\partial^{2}\log p\left(x;\boldsymbol{\theta}\right)}{\partial^{2}\mu}=-\frac{1}{\sigma^{2}}.
\end{align}
If we address the scenario in which $\mu$ and $\sigma^{2}$ are unknown parameters, then the classical multiparameter estimation strategy is required. Using equation (\ref{CFI}), it is easy to verify that
\begin{align}
	\left[\mathbb{F}\left(\boldsymbol{\boldsymbol{ \theta}} \right)\right]_{\mu\mu}&=\int p\left(\boldsymbol{x};\boldsymbol{\boldsymbol{ \theta}}\right)\left(\frac{\partial\log p\left(\boldsymbol{x};\boldsymbol{\boldsymbol{ \theta}}\right)}{\partial\mu}\right)^{2}dx\notag\\&=\int\frac{1}{p\left(\boldsymbol{x};\boldsymbol{\boldsymbol{ \theta}}\right)}\left(\frac{\partial p\left(\boldsymbol{x};\boldsymbol{\boldsymbol{ \theta}}\right)}{\partial\mu} \right)^{2}dx\notag\\&=-\int p\left(\boldsymbol{x};\boldsymbol{\boldsymbol{\theta}}\right)\frac{\partial^{2} \log p\left(\boldsymbol{x};\boldsymbol{\boldsymbol{ \theta}}\right)}{\partial^{2}\mu}dx\notag\\&=-\textbf{E}\left[\frac{\partial^{2}\log p\left(\boldsymbol{x};\boldsymbol{\boldsymbol{ \theta}}\right)}{\partial^{2}\mu}\right].
\end{align}
Thus, we calculate all the components of the classical Fisher information matrix in the same way, and eventually we find
\begin{equation*}
	\mathbb{F}\left(\boldsymbol{\boldsymbol{ \theta}} \right)_{N\left(\mu,\sigma\right)}= 
	\begin{pmatrix}
		\left[\mathbb{F}\left(\boldsymbol{\boldsymbol{ \theta}} \right)\right]_{\mu\mu} & \left[\mathbb{F}\left(\boldsymbol{\boldsymbol{ \theta}} \right)\right]_{\mu\sigma^{2}} \\
		\left[\mathbb{F}\left(\boldsymbol{\boldsymbol{ \theta}} \right)\right]_{\sigma^{2}\mu} & \left[\mathbb{F}\left(\boldsymbol{\boldsymbol{ \theta}} \right)\right]_{\sigma^{2}\sigma^{2}}
	\end{pmatrix}=\begin{pmatrix}
-\textbf{E}\left[\frac{\partial^{2}\log p\left(\boldsymbol{x};\boldsymbol{\boldsymbol{ \theta}}\right)}{\partial^{2}\mu}\right] & -\textbf{E}\left[\frac{\partial^{2}\log p\left(\boldsymbol{x};\boldsymbol{\boldsymbol{ \theta}}\right)}{\partial\mu\partial\sigma^{2}}\right] \\
	-\textbf{E}\left[\frac{\partial^{2}\log p\left(\boldsymbol{x};\boldsymbol{\boldsymbol{ \theta}}\right)}{\partial\sigma^{2}\partial\mu}\right] &-\textbf{E}\left[\frac{\partial^{2}\log p\left(\boldsymbol{x};\boldsymbol{\boldsymbol{ \theta}}\right)}{\partial^{2}\sigma^{2}}\right]
\end{pmatrix}=\begin{pmatrix}
\frac{1}{\sigma^{2}}&0\\
0&\frac{1}{2\sigma^{4}}
\end{pmatrix}.
\end{equation*}
At this step, we have learned that the Fisher information depends not only on the properties of the system but also on the process that causes its evolution and on the chosen measurement due to the probability law that gives us the expression of $p\left(\boldsymbol{x};\boldsymbol{\boldsymbol{\theta}}\right)$ (see Eq.\ref{FC}). Then, one of the goals of metrology being to maximize the Fisher information, it is therefore necessary to determine which is the best protocol for measuring the system. For this reason, it would be good to know what is the best precision achievable if we use the optimal measurement protocol, even if we do not know what it consists of. We will therefore extend the classical discussion we have had so far to quantum metrology by introducing the notion of quantum Fisher information and quantum estimation theory.
\section{Quantum estimation theory}
The universal probabilistic nature of quantum mechanics leads us to work regularly with probabilities. Moreover, if we investigate areas related to experiments or physical realizations, this probabilistic nature of quantum mechanics becomes even more visible. Furthermore, some exotic features such as entanglement arise from quantum mechanics, which are directly related to the probabilistic properties of the quantum system. In the quantum estimation framework, the background of the parameter estimation problem remains virtually unchanged from the classical framework. Here, an unknown parameter $\theta$ encoded in a quantum state $\rho_{\theta}$ builds a quantum statistical model, then individual qubits can be measured using POVM measurements and these results $\left\lbrace \theta_{1},...,\theta_{n}\right\rbrace$ are used for the construction of an estimate $\boldsymbol{ \theta}$. The major distinction from the classical case is that the probability density function $p\left(x;\boldsymbol{ \theta}\right)$ is not arbitrary, but is a function that depends on both the quantum state $\rho$ and the applied measurement. Moreover, the results of the POVM measurements $\left\lbrace \theta_{1},...,\theta_{n}\right\rbrace$, analogous to $\left\lbrace x_{1},...,x_{n}\right\rbrace$ in the classical framework, are not necessarily independent of each other due to the quantum correlations existing in the probe quantum state. This allows estimates to be made with a super-classical precision known as the Heisenberg limit \cite{Pezze2009,Paris2009}.\par

The purpose of this section is to describe quantum systems from the point of view of metrology. We present a complete review of the principal theoretical tools needed to address the parameters estimation problem via quantum measures and quantum states, starting from the original wording developed by Holevo \cite{Holevo1973,Holevo2011,Hayashi2005} and Helstrom \cite{Helstrom1967,Helstrom1976}, up to the most recent analytical results concerning the two estimation strategies.

\subsection{Quantum statistical schemes}
By analogy with the classical case, a quantum statistical model $S_{Q}$ is defined by a family of density operators in a Hilbert space $\cal{H}$ and is parameterized by $n$ real parameters $\theta\in \Theta\subset\mathbb{R}^{n}$, where $\Theta$ is an estimated parameter space. This means that
\begin{equation}
	S_{Q}=\left\lbrace\rho_{\theta}=\Lambda_{\theta}\left(\rho\right): \hspace{0.5cm}\theta=\left(\theta^{1},\theta^{2},...,\theta^{n}\right)\in\Theta\right\rbrace, 
\end{equation}
where the application $\Lambda_{\theta}$ is injective (i.e., there is a reverse of $\Lambda_{\theta}$). In fact, a quantum statistical model typically looks like this: The considered system $S$ is prepared at time $t=0$ in an initial state $\rho\equiv\rho\left(t=0\right)$, and then passes through a quantum channel $\Lambda_{\theta }$ which depends on the true value of one or more parameters $\theta$. So the associated model is defined as $\rho_{\theta}:=\Lambda_{\theta}\left(\rho\right)$. A typical example of application $\Lambda_{\theta }$ is the unit channel generated by the Hamiltonian of the system $S$, namely $\rho_{\theta}=U_{t}\rho U_{t}^{\dagger}$, and $U_{t}=\exp\left(-itH_{\theta}\right)$ where the parameter $\theta$ is called the Hamiltonian parameter.\par 
As an illustration, we consider here the thermal model which describes the equilibrium state of a quantum system interacting with a thermal reservoir, with
\begin{equation}
	\rho_{\beta}=\frac{\exp\left(-\beta H\right)}{Z}, \hspace{1cm}{\rm and}\hspace{1cm} Z={\rm Tr}\left(\exp\left(-\beta H\right)\right),
\end{equation}
where the interest parameter, conventionally referred to as $\beta$,
is the inverse of the reservoir temperature and $H$ is the Hamiltonian of the system.\par
Usually, any quantum statistical model $S=\left\lbrace \rho_{\boldsymbol{\theta}}\right\rbrace_{\boldsymbol{\theta}\in\Theta}$, performing a measurement with the operators $\left\lbrace \Pi_{x}\right\rbrace_{x\in X}$, leads to a classical statistical model via the formula $p_{\theta}\left(x \right)=p\left(x;\theta\right)={\rm Tr}\left(\rho_{\theta}\Pi_{x} \right)$, where the sample space $X$ is expected to be countable. To illustrate, let us take a state of a two-level quantum system that we can decompose on the eigenstate basis $\left\lbrace\left|0 \right\rangle,\left|1\right\rangle \right\rbrace$ in terms of two angles $\left\lbrace\theta^{1}\equiv\theta,\theta^{2}\equiv\varphi\right\rbrace$ as follows 
\begin{equation}
	\left|\psi \right\rangle=\cos\left(\frac{\theta}{2} \right)\left|0 \right\rangle+e^{i\varphi}\sin\left(\frac{\theta}{2} \right)\left|1 \right\rangle,
\end{equation}
with $0\leq\theta\leq\pi$, $0\leq\varphi\leq2\pi$ and evolving in a two dimensional Hilbert space. The state of the qubit is described by the Bloch vector $\vec{r}$ as shown below:
\begin{equation}
	\hat{\rho}=\frac{1}{2}\sum_{i=0}^{3}r_{i}\hat{\sigma_{i}}, \hspace{1cm}\hat{\sigma_{0}}=\hat{\openone}_{2\times2}, \hspace{1cm}r_{i}={\rm Tr}\left(\hat{\sigma_{i}}\hat{\rho}\right),\label{4.6}
\end{equation}
where $\sigma_{i}$ is the Pauli operator in the direction $i$ ($i = x, y, z$). Then, the state of a two-level system is visualized on the Bloch sphere by the Bloch vector
\begin{equation}
	\vec{r}=\left( {\begin{array}{*{20}{c}}
			\left\langle {\hat{\sigma_{x}}}\right\rangle \\
			\left\langle {\hat{\sigma_{y}}}\right\rangle\\
			\left\langle {\hat{\sigma_{z}}}\right\rangle\end{array}} \right)=\left( {\begin{array}{*{20}{c}}
			\cos\theta\sin\varphi \\
			\sin\theta\sin\varphi\\
			\cos\varphi\end{array}} \right).
\end{equation}
If the state $\hat{\rho}$ is a pure state, the density matrix (\ref{4.6}) reduces to
\begin{equation}
	\hat{\rho}=\frac{1}{2}\left( {\begin{array}{*{20}{c}}
			1+\cos\frac{\theta}{2}&1-i\cos\frac{\theta}{2}\sin\frac{\theta}{2}\\
			1+i\cos\frac{\theta}{2}\sin\frac{\theta}{2}&1-\cos\frac{\theta}{2}\end{array}} \right).\label{4.8}
\end{equation}
Assuming that we perform the projection value measurement (PVM) of the operator $\sigma_{z}$ on the state (\ref{4.8}), which is described by the set of projection operators
\begin{equation}
	\hat{\Pi}_{+}\equiv\frac{1}{2}\left(\hat{\openone}+ \hat{\sigma_{z}}\right),\hspace{1cm}{\rm and}\hspace{1cm} \hat{\Pi}_{-}\equiv\frac{1}{2}\left(\hat{\openone}- \hat{\sigma_{z}}\right),
\end{equation}
then the probability distribution of the result is obtained by the following calculation
\begin{equation}
	p\left(+,\theta \right)\equiv{\rm Tr} \left(\hat{\Pi}_{+}\hat{\rho}\right)=\frac{1}{2}\left(1+\cos\frac{\theta}{2} \right),\hspace{1cm} p\left(-,\theta\right)\equiv{\rm Tr} \left(\hat{\Pi}_{-}\hat{\rho}\right)=\frac{1}{2}\left(1-\cos\frac{\theta}{2}\right).
\end{equation}
Consequently, this statistical model is parameterized by
\begin{equation}
	x=\left\lbrace +,-\right\rbrace,\hspace{1cm} p\left(\pm,\theta \right)=\frac{1}{2}\left(1\pm\cos\frac{\theta}{2} \right), 
\end{equation}
and the parameter space $\Theta$ becomes
\begin{equation}
	\Theta=\left\lbrace\boldsymbol{\theta}=\left(\theta,\varphi\right),\hspace{0.5cm}0\leq\theta\leq\pi, \hspace{0.5cm} 0\leq\varphi\leq2\pi, \hspace{0.5cm}n=2 \right\rbrace.
\end{equation}

\subsection{Quantum Fisher Information}
The Fisher information defines a limit on the precision of unknown parameters related to a specific measurement strategy. It is obvious that different strategies extract different amounts of information. Typically, by employing a particular measurement scheme that extracts information about a given characteristic, we exclude the possibility of measuring complementary observables in the same experiment. For a quantum system, the variation of the unknown parameter $\theta$ in the system state $\rho_{\theta}$ provides an upper quantum limit on the classical Fisher information, ${\cal F}\left( p\left(x;\theta\right)\right)\leq{\cal F}_{Q}\left(\rho_{\theta}\right)$, with the quantum Fisher information ${\cal F}_{Q}\left(\rho_{\theta}\right)$ is given by
\begin{equation}
	{\cal F}_{Q}\left(\rho_{\theta}\right)={\rm Tr}\left(\hat{L}_{\theta}^{2}\rho_{\theta}\right), \label{QF}
\end{equation}
where the symmetric logarithmic derivative $\hat{L}_{\theta}$ is a self-adjoint operator that relates to the variation of the state $\rho_{\theta}$ with infinitesimal changes in the value of the unknown parameter $\theta$. This is determined by solving the following Lyapunov equation
\begin{equation}
	\frac{\partial\rho_{\theta}}{\partial\theta}=\frac{1}{2}\left(\hat{L}_{\theta}\rho_{\theta}+\rho_{\theta}\hat{L}_{\theta} \right).\label{DLS}
\end{equation}
In other words, the quantum Fisher information (\ref{QF}) can be formally derived by maximizing the classical Fisher information (\ref{FC}) over all possible measurements that can be made on the state $\rho$, and qualitatively, is a measure of the amount of information that a state contains about the parameter $\theta$. A general solution for the symmetric logarithmic derivative $L_{\theta}$ can be found as
\begin{equation}
	\hat{L}_{\theta}=2\int_{0}^{\infty}\exp\left[-\rho_{\theta}t \right]\frac{\partial\rho_{\theta}}{\partial\theta}\exp\left[-\rho_{\theta}t \right]dt,
\end{equation}
which is not an easy equation to handle. We will derive here an alternative form explicitly in terms of eigenvalues $\lambda_{i}$ and eigenvectors $\left|\vartheta_{i} \right\rangle$ of the density matrix $\rho_{\theta}$. By writing $\rho_{\theta}$ in its diagonal basis; $\rho_{\theta}\left|\vartheta_{i} \right\rangle=\lambda_{i}\left|\vartheta_{i} \right\rangle$. Then we can write
\begin{align}
	\left(\frac{\partial\rho_{\theta}}{\partial\theta} \right)_{ij}&=\left\langle\vartheta_{i}\right|\frac{\partial\rho_{\theta}}{\partial\theta}\left|\vartheta_{j} \right\rangle\notag\\&=\frac{1}{2}\left[\left\langle\vartheta_{i}\right|\hat{L}_{\theta}\rho_{\theta}\left|\vartheta_{j} \right\rangle\notag+\left\langle\vartheta_{i}\right|\rho_{\theta}\hat{L}_{\theta}\left|\vartheta_{j} \right\rangle\notag \right]\notag\\&=\frac{1}{2}\left[\lambda_{j}\left(\hat{L}_{\theta}\right)_{ij}+\lambda_{i}\left(\hat{L}_{\theta}\right)_{ij}\right].  
\end{align}
Next, we solve to find $L_{\theta}$ as
\begin{equation}
	\left(\hat{L}_{\theta}\right)_{ij}=\frac{2}{\lambda_{i}+\lambda_{j}}\left\langle\vartheta_{i}\right|\frac{\partial\rho_{\theta}}{\partial\theta}\left| \vartheta_{j}\right\rangle,\label{dls}
\end{equation}
where the denominator includes only those terms that satisfy $\lambda_{i}+\lambda_{j}\neq0$. 

\subsubsection{\bf Analytical expression of QFI for pure states}
Here, we shall focus on the much simpler scenario in which our quantum state remains a pure state, such that $\rho_{\theta}^{2}=\rho_{\theta}=\left|\psi_{\theta} \right\rangle\left\langle\psi_{\theta}\right|$. Therefore, we have
\begin{equation}
	\frac{\partial\rho_{\theta}}{\partial\theta}=\frac{\partial\rho_{\theta}^{2}}{\partial\theta}=\rho_{\theta}\frac{\partial\rho_{\theta}}{\partial\theta}+\frac{\partial\rho_{\theta}}{\partial\theta}\rho_{\theta}.
\end{equation}
When we compare this with the equation (\ref{DLS}), we get
$\hat{L}_{\theta}=2\frac{\partial\rho_{\theta}}{\partial\theta}$. We therefore find
\begin{align}
	{\cal F}_{Q}\left(\rho_{\theta}\right)=4{\rm Tr}\left[\rho_{\theta}\left(\frac{\partial\rho_{\theta}}{\partial\theta}\right)^{2} \right]=4\left[\left\langle\hat{\psi_{\theta}}|\hat{\psi_{\theta}}\right\rangle - |\left\langle \hat{\psi_{\theta}}|\psi_{\theta}\right\rangle|^{2} \right],  
\end{align}
where $\left|\hat{\psi_{\theta}}\right\rangle=\partial_{\theta}\left|\psi_{\theta}\right\rangle$. The last expression, combined with the assumption of unitary transformations ${\cal U}_{\theta}\left|\psi_{0}\right\rangle\equiv\left|\psi_{\theta}\right\rangle$
where ${\cal U}_{\theta}=\exp\left[-i\theta \hat{H}_{\theta} \right]$ and with $\hat{H}_{\theta}$ is a hermitian operator, brings us to the last simplified form of quantum Fisher information. For these states, the analytical expression of the quantum Fisher information is expressed as
\begin{equation}
	{\cal F}_{Q}\left(\left|\psi_{\theta} \right\rangle\left\langle\psi_{\theta}\right|\right)\equiv{\cal F}_{Q}\left(\left|\psi_{0}\right\rangle,\hat{H}_{\theta}\right)=4{\rm Var}\left(\hat{H}_{\theta}\right),
\end{equation}
which is proportional to the variance of the Hermitian operator $\hat{H}_{\theta}$. Moreover, an important property of the quantum Fisher information is its convexity. This property implies that ${\cal F}_{Q}\left(\rho\right)$ achieves its maximum value on pure states and hence, the mixed quantum states cannot increase the achievable estimation sensitivity. We can therefore generalize the last formula as
\begin{equation}
	{\cal F}_{Q}\left(\rho\right)\leq4{\rm Var}\left(\hat{H}_{\theta}\right),
\end{equation}
with equality applied to the pure states.
\subsubsection{\bf Analytical expression of QFI for mixed states}
In this part, we provide a derivation of the quantum Fisher information for mixed quantum states. For mixed states, we can rewrite the state $\rho_{\theta}\equiv\rho$ as a decomposition $\rho=\sum_{i}p_{i}\left|\psi_{i}\right\rangle\left\langle \psi_{i}\right|$ where $p_{i}$ and $\left| \psi_{i}\right\rangle$ are respectively the eigenstates and the eigenvectors of the density matrix $\rho$. The quantum Fisher information in this case can generally be expressed as follows
\begin{align}
	{\cal F}_{Q}\left(\rho\right)=4\sum_{i=0}^{s}\left(\left(\partial_{\theta}\sqrt{p_{i}} \right)^{2}+ p_{i}\left\langle\partial_{\theta}\psi_{i}|\partial_{\theta}\psi_{i}\right\rangle-p_{i}|\left\langle\psi_{i}|\partial_{\theta}\psi_{i} \right\rangle|^{2}\right) -8\sum_{i\neq j}\frac{p_{i}p_{j}}{p_{i}+p_{j}}|\left\langle\psi_{i}|\partial_{\theta}\psi_{j} \right\rangle|^{2},
\end{align}
where the Hermitian operator $\hat{H}_{\theta}$ is defined by $\hat{H}_{\theta}=-i{\cal U}_{\theta}^{\dagger}\partial_{\theta}{\cal U}_{\theta}$. The symmetric logarithmic derivative (\ref{dls}) then becomes
\begin{equation}
	\hat{L}_{\theta}=2\sum_{ij}\frac{\left\langle \psi_{i}\right| \partial_{\theta}\rho\left| \psi_{j}\right\rangle }{p_{i}+p_{j}}\left|\psi_{i} \right\rangle\left\langle\psi_{j}\right|. 
\end{equation}
Subsequently, by superimposing the definition of the symmetric logarithmic derivative (\ref{DLS}) with two states of the eigenbasis, one gets
\begin{align}
	\left\langle \psi_{i}\right|\partial_{\theta}\rho\left| \psi_{j}\right\rangle&=\frac{1}{2}\left(\left\langle \psi_{i}\right|\hat{L}_{\theta}\rho\left| \psi_{j}\right\rangle+\left\langle \psi_{i}\right|\rho \hat{L}_{\theta}\left| \psi_{j}\right\rangle \right)\notag\\&=\frac{1}{2}\left(\left\langle \psi_{i}\right|\hat{L}_{\theta}\sum_{k}p_{k}\left|\psi_{k}\right\rangle\left\langle\psi_{k}\right| \left| \psi_{j}\right\rangle+\left\langle \psi_{i}\right|\sum_{k}p_{k}\left|\psi_{k}\right\rangle\left\langle\psi_{k}\right|\hat{L}_{\theta} \left| \psi_{j}\right\rangle \right)\notag\\&=\frac{1}{2}\left(\left\langle \psi_{i}\right|\hat{L}_{\theta}p_{j}\left|\psi_{j}\right\rangle+\left\langle \psi_{i}\right|p_{i}\hat{L}_{\theta}\left|\psi_{j}\right\rangle  \right)\notag\\&=\frac{1}{2}\left(p_{i}+p_{j} \right)\left(\hat{L}_{\theta}\right)_{ij},     
\end{align}
wherein we have determined $\left(\hat{L}_{\theta}\right)_{ij}=\left\langle \psi_{i}\right| \hat{L}_{\theta}\left|\psi_{j}\right\rangle$. Therefore, we can write the quantum Fisher information as
\begin{align}
	{\cal F}_{Q}\left(\rho\right)&=\sum_{i}\left\langle \psi_{i} \right| \hat{L}_{\theta}^{2}\rho\left|\psi_{i}\right\rangle\notag\\&=\sum_{ij}\left\langle \psi_{i} \right| \hat{L}_{\theta}\left|\psi_{j} \right\rangle\left\langle\psi_{j} \right| \hat{L}_{\theta}\rho\left|\psi_{i}\right\rangle\notag\\&=\sum_{ij}\left\langle \psi_{i} \right| \hat{L}_{\theta}\left|\psi_{j} \right\rangle\left\langle\psi_{j} \right| \hat{L}_{\theta}\left|\psi_{i}\right\rangle p_{i}\notag\\&=\sum_{ij}\left(\hat{L}_{\theta}\right)_{ij}\left(\hat{L}_{\theta}\right)_{ji}p_{i}.
\end{align}
Now, the symmetric logarithmic derivative $\left(\hat{L}_{\theta}\right)_{ij}$ can be rewritten as follows
\begin{equation}
	\left(\hat{L}_{\theta}\right)_{ij}=\frac{2\left\langle \psi_{i}\right|\partial_{\theta}\rho\left| \psi_{j}\right\rangle}{p_{i}+p_{j}}.
\end{equation}
Using this expression, the quantum Fisher information is expressed by the following formula
\begin{equation}
	{\cal F}_{Q}\left(\rho\right)=\sum_{i=0}^{s}\sum_{j=0}^{t}\frac{4p_{i}}{\left(p_{i}+p_{j}\right)^{2}}|\left\langle \psi_{i}\right|\partial_{\theta}\rho\left| \psi_{j}\right\rangle|^{2}.
\end{equation}
Here we have entered the two indices $s$ and $t$. The state supports only $s$, so all indices $p_{t}=0$ for $t\geq s$, because the operator $\partial_{\theta}\rho$ is Hermitian. Afterward, we can show from the spectral decomposition of $\rho$ that
\begin{align}
	\left\langle \psi_{i}\right|\partial_{\theta}\rho\left| \psi_{j}\right\rangle&=\sum_{k}\left(\left\langle \psi_{i}\right|\partial_{\theta}p_{k}\left| \psi_{k}\right\rangle\left\langle \psi_{k}| \psi_{j}\right\rangle+p_{k}\left\langle \psi_{i}| \partial_{\theta}\psi_{k}\right\rangle\left\langle \psi_{k}| \psi_{j}\right\rangle+p_{i} \left\langle \psi_{i}|\psi_{k}\right\rangle\left\langle \partial_{\theta}\psi_{k}| \psi_{j}\right\rangle\right)\notag\\&= \partial_{\theta}p_{i}\delta_{ij}+p_{j}\left\langle \psi_{i}| \partial_{\theta}\psi_{j}\right\rangle+p_{i}\left\langle \partial_{\theta}\psi_{i}| \psi_{j}\right\rangle.
\end{align}
We use the fact that the resolution of the identity becomes zero when it is differentiated, i.e.,
\begin{equation}
	\partial_{\theta}\openone\equiv\sum_{i}\left| \partial_{\theta}\psi_{i}\right\rangle\left\langle\psi_{i}\right|+\left|\psi_{i}\right\rangle\left\langle\partial_{\theta}\psi_{i}\right|=0,
\end{equation}
this implies $\left\langle\partial_{\theta}\psi_{i}|\psi_{j}\right\rangle=-\left\langle\psi_{i}|\partial_{\theta}\psi_{j}\right\rangle$. So we find that
\begin{equation}
	\left\langle \psi_{i}\right|\partial_{\theta}\rho\left| \psi_{j}\right\rangle=\partial_{\theta}p_{i}\delta_{ij}+\left(p_{j}-p_{i} \right)\left\langle\psi_{i}|\partial_{\theta}\psi_{j} \right\rangle.
\end{equation}
We now insert this into the above equation to find
\begin{align}
	{\cal F}_{Q}\left(\rho\right)&=\sum_{i=0}^{s}\sum_{j=0}^{t}\frac{4p_{i}}{\left(p_{i}+p_{j} \right)^{2}}|\partial_{\theta}p_{i}\delta_{ij}+\left(p_{j}-p_{i} \right)\left\langle\psi_{i}|\partial_{\theta}\psi_{j} \right\rangle|^{2}\notag\\&=\sum_{i=0}^{s}\frac{1}{p_{i}}\left(\partial_{\theta}p_{i} \right)^{2} +\sum_{i=0}^{s}\sum_{j=0}^{t}\frac{4p_{i}\left(p_{i}-p_{j} \right)^{2} }{\left(p_{i}+p_{j} \right)^{2}}|\left\langle\psi_{i}|\partial_{\theta}\psi_{j} \right\rangle|^{2}.
\end{align}
We notice that the second sum can be divided into two sums; one that runs on the support until $s$ and the other $j$ the support from $s+1$ to $t$. Hence
\begin{align}
	\sum_{i=0}^{s}\sum_{j=0}^{t}\frac{4p_{i}\left(p_{i}-p_{j} \right)^{2} }{\left(p_{i}+p_{j} \right)^{2}}|\left\langle\psi_{i}|\partial_{\theta}\psi_{j} \right\rangle|^{2}=\sum_{i,j=0}^{s}\frac{4p_{i}\left(p_{i}-p_{j} \right)^{2} }{\left(p_{i}+p_{j} \right)^{2}}|\left\langle\psi_{i}|\partial_{\theta}\psi_{j} \right\rangle|^{2}+\sum_{i=0}^{s}\sum_{j=s+1}^{t}4p_{i}|\left\langle\psi_{i}|\partial_{\theta}\psi_{j} \right\rangle|^{2}.
\end{align}
It should be noted that all $p_{s+1}=0$, thus the second sum can be written like this
\begin{equation}
	\sum_{j=s+1}^{t}\left|\psi_{j} \right\rangle \left\langle\psi_{j}\right|=\openone-\sum_{j=0}^{s}\left|\psi_{j} \right\rangle \left\langle\psi_{j}\right|.
\end{equation}
Putting this expression into the above expression gives the following result
\begin{align}
	\sum_{i=0}^{s}\sum_{j=s+1}^{t}4p_{i}|\left\langle\psi_{i}|\partial_{\theta}\psi_{j} \right\rangle|^{2}&=\sum_{i=0}^{s}\sum_{j=s+1}^{t}4p_{i}\left\langle\partial_{\theta}\psi_{i}|\psi_{j}\right\rangle\left\langle\psi_{j}|\partial_{\theta}\psi_{i} \right\rangle\notag\\&=\sum_{i=0}^{s}4p_{i}\left\langle \partial_{\theta}\psi_{i}\right|\left[\openone-\sum_{j=0}^{s}\left|\psi_{j} \right\rangle \left\langle\psi_{j}\right| \right] \left|\partial_{\theta}\psi_{i}\right\rangle\notag\\&= \sum_{i=0}^{s}4p_{i}\left\langle\partial_{\theta}\psi_{i}|\partial_{\theta}\psi_{i}\right\rangle-\sum_{i,j=0}^{s}4p_{i}|\left\langle\psi_{j}|\partial_{\theta}\psi_{i} \right\rangle|^{2}.
\end{align}
If we put all these elements together, we come up with
\begin{align}
	{\cal F}_{Q}\left(\rho\right)&=\sum_{i=0}^{s}\frac{1}{p_{i}}\left(\partial_{\theta}p_{i} \right)^{2} +\sum_{i,j=0}^{s}\frac{4p_{i}\left(p_{i}-p_{j} \right)^{2} }{\left(p_{i}+p_{j} \right)^{2}}|\left\langle\psi_{i}|\partial_{\theta}\psi_{j} \right\rangle|^{2}+\sum_{i=0}^{s}4p_{i}\left\langle\partial_{\theta}\psi_{i}|\partial_{\theta}\psi_{i}\right\rangle-\sum_{i,j=0}^{s}4p_{i} |\left\langle\psi_{j}|\partial_{\theta}\psi_{i} \right\rangle|^{2}\notag\\&=\sum_{i=0}^{s}\frac{1}{p_{i}}\left(\partial_{\theta}p_{i} \right)^{2}+\sum_{i=0}^{s}4p_{i}\left\langle\partial_{\theta}\psi_{i}|\partial_{\theta}\psi_{i}\right\rangle-8\sum_{i,j=0}^{s}\frac{p_{i}p_{j}}{p_{i}+p_{j}} |\left\langle\psi_{i}|\partial_{\theta}\psi_{j} \right\rangle|^{2}.\label{466}
\end{align}
Furthermore, we can also write the first part of this expression in the form
\begin{equation}
	\sum_{i=0}^{s}\frac{1}{p_{i}}\left(\partial_{\theta}p_{i} \right)^{2}=4\sum_{i=0}^{s}\left(\partial_{\theta}\sqrt{p_{i}} \right)^{2}.
\end{equation}
We now remark that the third sum of the equation (\ref{466}) holds both diagonal and non-diagonal values. As a result, this expression can be further simplified and the quantum Fisher information is finally expressed as:
\begin{align}
	{\cal F}_{Q}\left(\rho\right)&=4\sum_{i=0}^{s}\left(\partial_{\theta}\sqrt{p_{i}} \right)^{2}+\sum_{i=0}^{s}4p_{i}\left\langle\partial_{\theta}\psi_{i}|\partial_{\theta}\psi_{i}\right\rangle-8\left(\sum_{i}\delta_{ij}+\sum_{i\neq j} \right)\frac{p_{i}p_{j}}{p_{i}+p_{j}} |\left\langle\psi_{i}|\partial_{\theta}\psi_{j} \right\rangle|^{2}\notag\\&=4\sum_{i=0}^{s}\left(\left(\partial_{\theta}\sqrt{p_{i}} \right)^{2}+ p_{i}\left\langle\partial_{\theta}\psi_{i}|\partial_{\theta}\psi_{i}\right\rangle-p_{i}|\left\langle\psi_{i}|\partial_{\theta}\psi_{i} \right\rangle|^{2}\right) -8\sum_{i\neq j}\frac{p_{i}p_{j}}{p_{i}+p_{j}}|\left\langle\psi_{i}|\partial_{\theta}\psi_{j} \right\rangle|^{2}.
\end{align}
When the states  $\left|\psi_{i}\right\rangle$ are evolving over time, we obtain $\left|\psi_{i}\left(t\right)\right\rangle={\cal U}_{\theta}\left(t \right)\left|\psi_{i}\left(0\right)\right\rangle$, which involves $\partial_{\theta}\left|\psi_{i}\left(t\right)\right\rangle=\partial_{\theta}{\cal U}_{\theta}\left(t \right)\left|\psi_{i}\left(0\right)\right\rangle$. In order to simplify the handling of this expression, we define the generator $\hat{H}_{\theta}=-i{\cal U}_{\theta}^{\dagger}\partial_{\theta}{\cal U}_{\theta}$. By applying the identity property ${\cal U}_{\theta}^{\dagger}{\cal U}_{\theta}=\openone$, we observe that
\begin{equation}
	{\cal U}_{\theta}^{\dagger}\partial_{\theta}{\cal U}_{\theta}=-\left( \partial_{\theta}{\cal U}_{\theta}^{\dagger} \right){\cal U}_{\theta}.
\end{equation}
In addition, we have
\begin{equation}
	\hat{H}_{\theta}^{2}=-{\cal U}_{\theta}^{\dagger}\left(\partial_{\theta}{\cal U}_{\theta}\right){\cal U}_{\theta}^{\dagger}\partial_{\theta}{\cal U}_{\theta}=\left(\partial_{\theta}{\cal U}_{\theta}\right)^{\dagger}\partial_{\theta}{\cal U}_{\theta}.
\end{equation}
In the case of a unitary evolution operator ${\cal U}_{\theta}$, the classical contribution vanishes ($\partial_{\theta}\sqrt{p_{i}}=0$).  Thereby we proceed to write
\begin{align}
	{\cal F}_{Q}\left(\rho\right)=4\sum_{i=0}^{s}p_{i}&\left( \left\langle\psi_{i}\left(0 \right)\right|\hat{H}_{\theta}^{2}\left| \psi_{i}\left(0 \right)\right\rangle-\left\langle\psi_{i}\left(0 \right)\right|\hat{H}_{\theta}\left| \psi_{i}\left(0 \right)\right\rangle\right)\notag\\& -8\sum_{i\neq j}\frac{p_{i}p_{j}}{p_{i}+p_{j}}|\left\langle\psi_{i}\left(0 \right)\right|\hat{H}_{\theta}\left| \psi_{i}\left(0 \right)\right\rangle|^{2}.
\end{align}
The following table summarizes the properties of classical and quantum estimation and the relationship between them:
\begin{center}
	\begin{table}[h!]
		\centering
	\begin{tabular}{|l|c|r|}
		\hline \multicolumn{2}{c|}{Estimation Theory}  \\
		\hline  Classical Strategy & Quantum Strategy  \\
		\hline Probability density $p\left(x;\theta\right)$ & Density matrix $\rho_{\theta}=\Lambda_{\theta}\left(\rho_{0}\right)$  \\
		\hline  Score ${\cal L}_{\theta}\left(x\right) =\frac{\partial}{\partial \theta}\log p\left(x;\theta\right)$ & Symmetric logarithmic derivative $\hat{L}_{\theta}$  \\
		\hline  $\frac{\partial p\left(x;\theta\right)}{\partial\theta}=\frac{1}{2}\left( {\cal L}_{\theta}p\left(x;\theta\right)+p\left(x;\theta\right){\cal L}_{\theta} \right) $ & $\frac{\partial\rho_{\theta}}{\partial\theta}=\frac{1}{2}\left(\hat{L}_{\theta}\rho_{\theta}+\rho_{\theta}\hat{L}_{\theta} \right)$ \\
		\hline  ${\rm Var}\left({\cal L}_{\theta}\right)=\sum_{x}p\left(x;\theta\right){\cal L}_{\theta}^{2}={\cal F}\left(\theta\right)$ & ${\rm Var}\left(\hat{L}_{\theta}\right)={\rm Tr}\left(\rho\hat{L}_{\theta}^{2}\right)={\cal F}_{Q}\left(\rho\right)$ \\
		\hline  $\bar{{\cal L}_{\theta}}=\sum_{x}p\left(x;\theta\right){\cal L}_{\theta}=0$ &$\left\langle \hat{L}_{\theta}\right\rangle={\rm Tr}\left(\rho\hat{L}_{\theta}\right)=\frac{\partial}{\partial \theta}{\rm Tr}\left(\rho\right)=0$  \\
		\hline 
	\end{tabular}
	\caption{Basic properties of classical and quantum estimation theory}
\end{table}
\end{center}
Interestingly, there's always an optimal measure for any $\rho$ such that ${\cal F}\left(\theta\right)={\cal F}_{Q}\left(\rho_{\theta} \right)$. This optimal measure is given by the projection on the eigenstates $\left|l\right\rangle$ of $\hat{L}_{\theta}$. In this case, we have $p\left( l;\theta\right)=\left\langle l\right|\rho_{\theta}\left|l\right\rangle$. Then
	\begin{equation}
		\frac{\partial p\left( l;\theta\right)}{\partial\theta}=\left\langle l\right|\frac{\partial \rho_{\theta}}{\partial\theta}\left|l\right\rangle=l p\left( l;\theta\right).
	\end{equation}
Therefore, it is easy to verify that
	\begin{equation}
		{\cal F}\left(\theta\right)=\sum_{l}\frac{1}{p\left( l;\theta\right)}\left( \frac{\partial p\left( l;\theta\right)}{\partial \theta}\right)^{2}=\sum_{l}l^{2}p\left( l;\theta\right)={\rm Tr}\left( \rho_{\theta} \hat{L}_{\theta}^{2}\right)={\cal F}_{Q}\left(\rho_{\theta} \right).
	\end{equation}
\section{Multiparametric quantum metrology}
So far, we have considered only one parameter to be estimated. However, there are tasks for which it is important to estimate several parameters. These scenarios include, for example, the simultaneous estimation of several phases and the estimation of several spins pointing in different directions or, in general, of parameters corresponding to non-commutable unitary generators. The theory described in the previous sections can be naturally generalized to the multiparameter estimation strategy. To determine the accuracy in this case, we need to compute the Fisher quantum information matrix, denoted $\boldsymbol{ {\cal F}}_{Q}\left(\boldsymbol{\theta}\right)$, which describes the limits of distinguishing infinitely close quantum states $\rho_{\boldsymbol{\theta}}$ and $\rho_{\boldsymbol{\theta}+d\boldsymbol{\theta}}$ where $\boldsymbol{\theta}=\left\lbrace\theta_{1},..\theta_{\mu},\theta_{\nu},..\theta_{n} \right\rbrace$. In general, larger elements of this matrix predict better distinguishability, which leads to better accuracy in estimating the parameter vector $\boldsymbol{\theta}$.\par 
On the other hand, how to increase the quantum Fisher information matrix is a challenging topic in quantum metrology. In fact, the estimation of a single parameter plays an important role in many ways, due to the existence of an optimal final state containing a maximum amount of the quantum Fisher information. Realistic problems can usually involve multiple parameters because there is no optimal state in which the quantum Fisher information matrix is larger than the other states. This is related to the fact that the optimal measures for the estimation of different parameters are not necessarily commutable. In addition, the Cramér-Rao inequality is not always saturable because the measurements of different parameters may be incompatible. All these restrictions make the simultaneous estimation of several parameters an important task in quantum metrology. In the same context, techniques for computing the quantum Fisher information matrix have seen rapid development in various scenarios and models. However, there is no literature that summarizes these techniques in a structured way. Therefore, this section provides extensive techniques for calculating the matrix $\boldsymbol{ {\cal F}}_{Q}\left(\boldsymbol{\theta}\right)$ across a variety of scenarios.

\subsection{Matrix representation of the quantum Fisher information}
Let us start by considering the general case, i.e. a family of quantum states $\rho_{\boldsymbol{\theta}}$ which depend on a set of $n$ different parameters $\boldsymbol{\theta}=\left\lbrace \theta_{\mu}\right\rbrace$, $\mu=1,..,n$. For each parameter involved, we can derive the operators of the symmetric logarithmic derivative as follows
\begin{equation}
	\frac{\partial\rho_{\boldsymbol{\theta}}}{\partial\theta_{\mu}}=\frac{1}{2}\left(\hat{L}_{\theta_{\mu}}\rho_{\boldsymbol{\theta}}+\rho_{\boldsymbol{\theta}}\hat{L}_{\theta_{\mu}} \right).\label{DLSM}
\end{equation}
The components of the quantum Fisher information matrix are defined as \cite{Paris2009}
\begin{equation}
	\boldsymbol{ {\cal F}}_{\theta_{\mu}\theta_{\nu}}\left(\rho_{\boldsymbol{\theta}}\right):=\frac{1}{2}{\rm Tr}\left(\rho_{\boldsymbol{\theta}}\left\lbrace\hat{L}_{\theta_{\mu}},\hat{L}_{\theta_{\nu}} \right\rbrace\right),\label{QFIM}
\end{equation}
where $\left\lbrace,\right\rbrace$ stands for anti-commutation and $\hat{L}_{\theta_{\mu}}\left(\hat{L}_{\theta_{\nu}} \right)$ is the symmetric logarithmic derivative for the parameter $\theta_{\mu}\left(\theta_{\nu}\right)$, which is derived by equation (\ref{DLSM}). It is generally recognized that ${\rm Tr}\left(\rho_{\boldsymbol{\theta}}\hat{L}_{\theta_{\mu}}\right)$, therefore using the above equation (\ref{QFIM}), $\boldsymbol{ {\cal F}}_{\theta_{\mu}\theta_{\nu}}\left(\rho_{\boldsymbol{\theta}}\right)$ can also be expressed as below
\begin{equation}
	\boldsymbol{ {\cal F}}_{\theta_{\mu}\theta_{\nu}}\left(\rho_{\boldsymbol{\theta}}\right)={\rm Tr}\left(\hat{L}_{\theta_{\nu}}\partial_{\theta_{\mu}}\rho_{\boldsymbol{\theta}} \right)=-{\rm Tr}\left(\rho_{\boldsymbol{\theta}}\partial_{\theta_{\mu}} \hat{L}_{\theta_{\nu}}\right).
\end{equation}
According to equation (\ref{QFIM}), the diagonal entry of the matrix $\boldsymbol{ {\cal F}}\left(\rho_{\boldsymbol{\theta}}\right)$ reads
\begin{equation}
	\boldsymbol{ {\cal F}}_{\theta_{\mu}\theta_{\mu}}\left(\rho_{\boldsymbol{\theta}}\right)={\rm Tr}\left(\rho_{\boldsymbol{\theta}}\hat{L}_{\theta_{\mu}}^{2} \right),
\end{equation}
which is exactly the quantum Fisher information for the parameter $\theta_{\mu}$. Besides, some properties of the quantum Fisher information have been well presented above. Similarly, the quantum Fisher information matrix also has some powerful properties that have been widely applied in practice. We present these properties here as follows
\begin{description}
	\item[*] $\boldsymbol{ {\cal F}}\left(\rho_{\boldsymbol{\theta}}\right)$ is real symmetric, namely $\boldsymbol{ {\cal F}}_{\theta_{\mu}\theta_{\nu}}=\boldsymbol{ {\cal F}}_{\theta_{\nu}\theta_{\mu}}\in\mathbb{R}^{2}$.
	\item[*] $\boldsymbol{ {\cal F}}\left(\rho_{\boldsymbol{\theta}}\right)$ is positive semi-definite, i.e. $\boldsymbol{ {\cal F}}\left(\rho_{\boldsymbol{\theta}}\right)\geq0$. If $\boldsymbol{ {\cal F}}>0$, then $\left[\boldsymbol{ {\cal F}}^{-1}\right]_{\theta_{\mu}\theta_{\mu}}=1/\boldsymbol{ {\cal F}}_{\theta_{\mu}\theta_{\mu}}$ for any $\theta_{\mu}$.
	\item[*] $\boldsymbol{ {\cal F}}\left(\rho_{\boldsymbol{\theta}}\right)=\boldsymbol{ {\cal F}}\left(U\rho_{\boldsymbol{\theta}}U^{\dagger}\right)$ for a unitary operation $U$.
	\item[*] $\boldsymbol{ {\cal F}}$ is a monotonic function under a completely positive and trace preserving map $\Lambda$, this means that $\boldsymbol{ {\cal F}}\left(\Lambda\left( \rho\right) \right)\leq\boldsymbol{ {\cal F}}\left( \rho\right)$.
	\item[*] Convexity property: $\boldsymbol{ {\cal F}}\left(p\rho_{1}+\left(1-p\right)\rho_{2} \right)\leq p\boldsymbol{ {\cal F}}\left(\rho_{1}\right)+\left(1-p \right)\boldsymbol{ {\cal F}}\left(\rho_{2}\right)$ for $p\in\left[0,1\right]$.
\end{description}
\subsection{Analytical expression of the quantum Fisher information matrix}
Here we review the derivation techniques of the quantum Fisher information matrix and some analytical results for specific cases. The standard derivation of this matrix generally assumes that the rank of the density matrix is complete, meaning that all eigenvalues of the $\rho$ density matrix are positive. More precisely, when we write $\rho=\sum_{i}\lambda_{i}\left|\vartheta_{i}\right\rangle\left\langle \vartheta_{i}\right|$, where $\lambda_{i}$ and $\left|\vartheta_{i}\right\rangle$ are the corresponding eigenvalue and eigenstate, we generally assume that $\lambda_{i}>0$, for any $0\leq i\leq {\rm dim}\left(\rho \right)-1$. With this assumption, the quantum Fisher information matrix elements can be written as follows
\begin{equation}
	\boldsymbol{ {\cal F}}_{\theta_{\mu}\theta_{\nu}}=\sum_{i,j=0}^{d-1}\frac{2\textsf{Re}\left(\left\langle \vartheta_{i}\right|\partial_{\theta_{\mu}} \rho\left|\vartheta_{j} \right\rangle\left\langle \vartheta_{j}\right|\partial_{\theta_{\nu}} \rho\left|\vartheta_{i} \right\rangle\right) }{\lambda_{i}+\lambda_{j}},
\end{equation}
with $\textsf{Re}$ denotes the real part and $d$ is the dimension of the density matrix. One can easily see that if the density matrix is not of full rank, there may be divergent terms in the above equation. To extend it to general density matrices, we can remove the divergent terms as follows
\begin{equation}
	\boldsymbol{ {\cal F}}_{\theta_{\mu}\theta_{\nu}}=\sum_{i,j=0,\lambda_{i}+\lambda_{j}\neq0}^{d-1}\frac{2\textsf{Re}\left(\left\langle \vartheta_{i}\right|\partial_{\theta_{\mu}} \rho\left|\vartheta_{j} \right\rangle\left\langle \vartheta_{j}\right|\partial_{\theta_{\nu}} \rho\left|\vartheta_{i} \right\rangle\right) }{\lambda_{i}+\lambda_{j}}.
\end{equation}
By substituting the spectral decomposition of $\rho$ in the above equation, we can rewrite it as
\begin{align}
	\boldsymbol{{\cal F}}_{\theta_{\mu}\theta_{\nu}}&=\sum_{i=0}^{d-1}\frac{\left(\partial_{\theta_{\mu}}\lambda_{i} \right) \left(\partial_{\theta_{\nu}}\lambda_{i} \right)}{\lambda_{i}}+\sum_{i\neq j,\lambda_{i}+\lambda_{j}\neq j}^{d-1}\frac{2\left(\lambda_{i}-\lambda_{j}\right)^{2}}{\left(\lambda_{i}+\lambda_{j}\right)}\textsf{Re}\left(\left\langle \vartheta_{i}|\partial_{\theta_{\mu}}\vartheta_{j} \right\rangle \left\langle\partial_{\theta_{\nu}}\vartheta_{j}|\vartheta_{i}\right\rangle \right)\notag\\&= \sum_{i}\frac{\left(\partial_{\theta_{\mu}}\lambda_{i} \right) \left(\partial_{\theta_{\nu}}\lambda_{i} \right)}{\lambda_{i}}+\sum_{i}4\lambda_{i}\textsf{Re}\left(\left\langle \partial_{\theta_{\mu}}\vartheta_{i}|\partial_{\theta_{\nu}}\vartheta_{i} \right\rangle\right)+\sum_{i\neq j}\frac{8\lambda_{i}\lambda_{j}}{\lambda_{i}+\lambda_{j}}\left(\left\langle \partial_{\theta_{\mu}}\vartheta_{i}|\vartheta_{j} \right\rangle \left\langle\vartheta_{j}|\partial_{\theta_{\nu}}\vartheta_{i}\right\rangle \right).
\end{align}
Due to the relationship between the quantum Fisher information matrix and the quantum Fisher information, one can easily obtain
\begin{align}
	\boldsymbol{{\cal F}}_{\theta_{\mu}\theta_{\mu}}=&{\cal F}_{Q}\left(\rho_{\theta_{\mu}} \right)=\sum_{i}\frac{\left(\partial_{\theta_{\mu}}\lambda_{i} \right)^{2}}{\lambda_{i}}+\sum_{i}4\lambda_{i}\left(\left\langle \partial_{\theta_{\mu}}\vartheta_{i}|\partial_{\theta_{\mu}}\vartheta_{i} \right\rangle\right)\notag\\&-\sum_{i\neq j}\frac{8\lambda_{i}\lambda_{j}}{\lambda_{i}+\lambda_{j}}|\left\langle \partial_{\theta_{\mu}}\vartheta_{i}|\vartheta_{j} \right\rangle|^{2}.
\end{align}
The Bloch representation is another tool also well used in quantum estimation theory. For a $d$-dimensional density matrix, it can be expressed by
\begin{equation}
	\rho=\frac{1}{d}\left(\openone+\sqrt{\frac{d\left(d-1\right)}{2}}\vec{r}.\vec{K} \right), 
\end{equation}
where $\vec{r}=\left(r_{1},r_{2},..,r_{m},..\right)^{T}$ is the Bloch vector and $\vec{K}$ is a ($d^{2}-1$)-dimensional vector of the generator of $SU\left(d \right)$ which satisfies ${\rm Tr}\left(K_{i}\right)=0$. The anti-commutation relation for them is $\left\lbrace K_{i},K_{j} \right\rbrace=\frac{4}{d}\delta_{ij}\openone+\sum_{m=1}^{d^{2}-1}v_{ijm}K_{m}$, and the commutation relation being $\left[K_{i},K_{j} \right]=i\sum_{m=1}^{d^{2}-1}\epsilon_{ijm}K_{m}$, where $v_{ijm}$ and $\epsilon_{ijm}$ are the symmetric and antisymmetric structure constants.\par
Very recently, Watanabe and his co-workers provided a simple analytical formula of the quantum Fisher information matrix for a general Bloch vector by considering the Bloch vector itself as parameters to be estimated. Using the Bloch representation of a $d$-dimensional density matrix, the quantum Fisher information matrix can be expressed as follows \cite{Watanabe2014}  
\begin{equation}
	\boldsymbol{{\cal F}}_{\theta_{\mu}\theta_{\nu}}=\left(\partial_{\theta_{\nu}}\vec{r} \right)^{T}\left(\frac{d}{2\left(d-1\right)}G-\vec{r}\vec{r}^{T} \right)\partial_{\theta_{\mu}}\vec{r}, 
\end{equation}
with $G$ is a real symmetric matrix whose components are
\begin{equation}
	G_{ij}=\frac{1}{2}{\rm Tr}\left(\rho\left\lbrace K_{i},K_{j} \right\rbrace \right)=\frac{2}{d}\delta_{ij}+\sqrt{\frac{d-1}{2d}}\sum_{m}v_{ijm}r_{m}.
\end{equation}
In an analytical point of view, the most widely used scenario of this theorem concerns one-qubit systems, where $\rho=\left(\openone+\vec{r}.\vec{\sigma}\right)$ with the vector of Pauli matrices $\vec{\sigma}=\left(\sigma_{x},\sigma_{y},\sigma_{z}\right)$. In such circumstances, the quantum Fisher information matrix reduces to
\begin{equation}
	\boldsymbol{{\cal F}}_{\theta_{\mu}\theta_{\nu}}=\left(\partial_{\theta_{\mu}}\vec{r} \right)\left(\partial_{\theta_{\nu}}\vec{r} \right)+\frac{\left(\vec{r}.\partial_{\theta_{\mu}}\vec{r} \right) \left(\vec{r}.\partial_{\theta_{\nu}}\vec{r} \right) }{1-|\vec{r}|^{2}},
\end{equation}
where $|\vec{r}|$ is the norm of $\vec{r}$. For a single qubit pure state, we find $\boldsymbol{{\cal F}}_{\theta_{\mu}\theta_{\nu}}=\left(\partial_{\theta_{\mu}}\vec{r} \right)\left(\partial_{\theta_{\nu}}\vec{r} \right)$.\par
For a general two-qubit state, the derivation of the quantum Fisher information matrix requires the diagonalization of a $4\times4$ density matrix, which is difficult to solve analytically. Nevertheless, some special two-qubit states, such as the $X$-state, can be analytically diagonalized. Indeed, the two-qubit state $\rho_{X}$ can be rewritten as a block diagonal in the form $\rho_{X}=\rho^{(0)}\oplus\rho^{(1)}$, where $\oplus$ represents the direct sum and
\begin{equation}
	\rho^{(0)}=\left( {\begin{array}{*{20}{c}}
			\rho_{11}&\rho_{14}\\
			\rho_{41}&\rho_{44}\end{array}} \right),\hspace{1cm}{\rm and}\hspace{1cm}\rho^{(1)}=\left( {\begin{array}{*{20}{c}}
			\rho_{22}&\rho_{23}\\
			\rho_{32}&\rho_{33}\end{array}} \right).
\end{equation}
It is worth noting that $\rho^{(0)}$ and $\rho^{(1)}$ are not density matrices because their trace are not normalized. The quantum Fisher information matrix for this state can be written as $\boldsymbol{{\cal F}}_{\theta_{\mu}\theta_{\nu}}=\boldsymbol{{\cal F}}_{\theta_{\mu}\theta_{\nu}}^{(0)}+\boldsymbol{{\cal F}}_{\theta_{\mu}\theta_{\nu}}^{(1)}$, where $\boldsymbol{{\cal F}}_{\theta_{\mu}\theta_{\nu}}^{(0)}(\boldsymbol{{\cal F}}_{\theta_{\mu}\theta_{\nu}}^{(1)})$ is the quantum Fisher information matrix for $\rho^{(0)}(\rho^{(1)})$. The eigenvalues of $\rho^{(i)}$ are
\begin{equation}
	\lambda_{\pm}^{(i)}=\frac{1}{2}\left({\rm Tr}\rho^{(i)}\pm\sqrt{\left({\rm Tr}\rho^{(i)}\right)^{2}-4{\rm det}\rho^{(i)} } \right), 
\end{equation}
and the corresponding eigenstates are
\begin{equation}
	\left|\vartheta_{\pm}^{(i)} \right\rangle={\cal N}_{\pm}^{(i)}\left(\frac{1}{2{\rm Tr}\left(\rho^{(i)}\sigma_{+} \right) }\left[{\rm Tr}\left(\rho^{(i)}\sigma_{+} \right)\pm \sqrt{\left({\rm Tr}\rho^{(i)}\right)^{2} -4{\rm det}\rho^{i}} \right],1\right)^{T},
\end{equation}
with ${\cal N}_{\pm}^{(i)}$ ($i=0,1$) is the normalization coefficient. In this case, the specific form of $\sigma_{z}$ et $\sigma_{+}$ are
\begin{equation}
	\sigma_{z}=\left( {\begin{array}{*{20}{c}}
			1&0\\
			0&-1\end{array}} \right),\hspace{2cm}\sigma_{+}=\left( {\begin{array}{*{20}{c}}
			0&1\\
			0&0\end{array}}\right).
\end{equation}
Based on the above mentioned information, $\boldsymbol{{\cal F}}_{\theta_{\mu}\theta_{\nu}}^{(i)}$ can be written specifically as follows
\begin{align}
	\boldsymbol{{\cal F}}_{\theta_{\mu}\theta_{\nu}}^{(i)}=&\sum_{k=\pm}\frac{\left(\partial_{\theta_{\mu}}\lambda_{k}^{(i)} \right)\left(\partial_{\theta_{\nu}}\lambda_{k}^{(i)} \right)}{\lambda_{k}^{(i)}}+\lambda_{k}^{(i)}\boldsymbol{{\cal F}}_{\theta_{\mu}\theta_{\nu}}\left(\left|\vartheta_{k}^{(i)} \right\rangle \right)\notag\\&-\frac{16{\det\rho^{(i)}}}{{\rm Tr}\rho^{(i)}}\textsf{Re}\left(\left\langle\partial_{\theta_{\mu}}\vartheta_{+}^{(i)}| \vartheta_{-}^{(i)}\right\rangle\left\langle\vartheta_{-}^{(i)}|\partial_{\theta_{\nu}} \vartheta_{+}^{(i)}\right\rangle\right),  
\end{align}
where $\boldsymbol{{\cal F}}_{\theta_{\mu}\theta_{\nu}}\left(\left|\vartheta_{k}^{(i)} \right\rangle \right)$ is the quantum Fisher information matrix for the state $\left|\vartheta_{k}^{(i)} \right\rangle$.
\subsection{Quantum Fisher Information Matrix via Vectorization Method}
Unlike the previous analytical expression of the quantum Fisher information matrix which required the diagonalization of the density matrix, Šafránek \cite{Safranek2018} has recently provided another analytical expression using the density matrix in the Liouville space formalism. Technically, by employing the vectorization of the density matrix $\rho$, the Lyapunov equations (\ref{DLSM}) are converted to a linear system that can be solved without diagonalizing the matrix $\rho$. This approach has proven to be successful and can be used for any finite dimensional density matrix. This is based on using the vec-operator which allows us to transform a density matrix into a column vector. Indeed, the operator ${\rm vec}\left[A\right]$ for any matrix $A\in \emph{M}^{n\times n}$ in the Liouville space, where $\emph{M}^{n\times n}$ refers to the space of $n\times n$ martrices, is given by
\begin{equation}
	{\rm vec}\left[A \right]=\left(a_{11},...,a_{n1},a_{12},...,a_{n2},a_{1n},...,a_{nn} \right)^{T}.
\end{equation}
Moreover, through the expression $A=\sum_{k,l=1}^{n}a_{kl}\left|k\right\rangle\left\langle l\right|$, the ${\rm vec}$-operator becomes
\begin{equation}
	{\rm vec}\left[A\right]=\left(\openone_{n\times n}\otimes A\right)\sum_{i=1}^{n}e_{i}\otimes e_{i}. 
\end{equation}
where $e_{i}$ stands for the computational components of $\emph{M}^{n\times n}$. Taking advantage of the Kronecker product properties \cite{Schacke2004}, we can find the results
\begin{align}
	{\rm vec}\left[AB\right]&=\left(\openone_{n\times n}\otimes A \right){\rm vec}\left[B\right]=\left(B^{T}\otimes\openone_{n\times n} \right){\rm vec} \left[A\right],\notag\\& {\rm Tr}\left(A^{T}B \right)={\rm vec}\left[A\right]^{\dagger} {\rm vec}\left[B\right],\notag\\&{\rm vec}\left[AXB\right]=\left(B^{\dagger}\otimes A\right){\rm vec}\left[X\right], \label{vecAXB}
\end{align}
that apply to any matrix $A$, $B$ and $X$. With the above properties (\ref{vecAXB}), it is straight forward to verify that the quantum Fisher information matrix can be expressed as follows
\begin{equation}
	\boldsymbol{{\cal F}}_{\theta_{\mu}\theta_{\nu}}=2{\rm vec}\left[\partial_{\theta_{\mu}}\rho\right]^{\dagger}\left(\rho\otimes\openone+\openone\otimes\rho^{*} \right)^{-1}{\rm vec}\left[\partial_{\theta_{\nu}}\rho \right],
\end{equation}
with $\rho^{*}$ being the complex conjugate of $\rho$. The symmetric logarithmic derivative operators in Liouville space, denoted by ${\rm vec}\left[\hat{L}_{\theta_{\mu}}\right]$, are reduced to
\begin{equation}
	{\rm vec}\left[\hat{L}_{\theta_{\mu}}\right]=2\left(\rho\otimes\openone+\openone\otimes\rho^{*} \right)^{-1}{\rm vec}\left[\partial_{\theta_{\mu}}\rho\right].
\end{equation}
The advantage of this method would be that it can be computed analytically for an arbitrary system and is only based on the computation of the inverse of the matrix $\left(\rho\otimes\openone+\openone\otimes\rho^{*} \right)$. For high-dimensional systems, efficient methods such as Cholesky decomposition may be required \cite{Krishnamoorthy2013}. It is worth noting that the above formula is only applicable if the density matrix $\rho$ has full rank and thus is invertible. If $\rho$ is not invertible, Šafránek gives another formal solution of quantum Fisher information matrix by using the invertible matrix $\rho_{s}\equiv\left(1-s\right)\rho+\frac{s}{d}\openone$ corresponding to $\rho$ such that the equation (\ref{QFIM}) becomes
\begin{equation}
	\boldsymbol{{\cal F}}_{\theta_{\mu}\theta_{\nu}}=2\lim_{s\rightarrow0}{\rm vec}\left[\partial_{\theta_{\mu}}\rho_{s}\right]^{\dagger}\left(\rho_{s}\otimes\openone+\openone\otimes\rho_{s}^{*} \right)^{-1}{\rm vec}\left[\partial_{\theta_{\nu}}\rho_{s} \right].
\end{equation}
To illustrate how the derived expressions can be used, we take the example of an isotropic Heisenberg $XY$ model submitted to an external magnetic field. Here, we estimated the temperature and the magnetic field of this quantum Heisenberg model in which the Hamiltonian describing the system can be written as \cite{Korepin1993,Kamta2002,Bakmou2019,Wang2002}
\begin{equation}
H=\sum_{n=1}^{2}J\left(S_{n}^{x}S_{n+1}^{x}+S_{n}^{y}S_{n+1}^{y} \right)+B\sum_{n=1}^{2}S_{n}^{z},\label{H}
\end{equation}
where $S_{n}^{i}=\frac{1}{2}\sigma_{n}^{i}$ (with $i=x,y,z$) and $\sigma_{n}^{i}$ are the spin-$\frac{1}{2}$ operators and the familiar Pauli matrices at site $n$ respectively, $B$ is the constant external magnetic field acting on the spin-$\frac{1}{2}$ particle oriented in the $z$-direction, and the parameter $J$ denotes the linear strength of the Heisenberg coupling coefficient for the spin interaction. Notice that the Heisenberg spin chain is antiferromagnetic for $J$ positive, and becomes ferromagnetic for $J$ negative. 
In terms of the raising and lowering operator $\sigma_{n}^{\pm}=\sigma_{n}^{x}\pm\sigma_{n}^{y}$, the Hamiltonian (\ref{H}) becomes
\begin{equation}
H=J\left(\sigma_{1}^{+}\sigma_{2}^{-} +\sigma_{2}^{+}\sigma_{1}^{-} \right)+\frac{B}{2}\left(\sigma_{1}^{z}+\sigma_{2}^{z}\right),  
\end{equation}
and the eigenvalues and eigenvectors associated to the Hamiltonian are easily obtained as
\begin{equation}
H\left|00\right\rangle=B\left|00\right\rangle,\hspace{1cm} H\left|11\right\rangle=-B\left|11\right\rangle,\hspace{1cm} {\rm and}\hspace{1cm} H\left|\psi^{\pm}\right\rangle=B\left|\psi^{\pm}\right\rangle,
\end{equation}
 where $\left|\psi^{\pm}\right\rangle=\left(\left|00\right\rangle\pm\left|11\right\rangle \right)/\sqrt{2}$ are the Bell maximally entangled states. For a system in equilibrium at temperature $T$, the state is described by density operator $\rho\left(T\right)=\exp\left(-\beta H\right)/Z$, where $Z={\rm Tr}\left[\exp\left(-\beta H\right)\right]$ with $\beta=1/k_{B}t$ and $k_{B}$ is the Boltzmann's constant. In the standard basis $\left\lbrace \left|00 \right\rangle,\left|01 \right\rangle,\left|10 \right\rangle,\left|11 \right\rangle\right\rbrace$, the resulting density matrix of the system can be written as
\begin{equation}
\rho\left(T\right)=\frac{1}{2\left(\cosh\left(\beta B \right)+\cosh\left(\beta J \right)\right)}\left( {\begin{array}{*{20}{c}}
		\exp\left(-\beta B \right) &0&0&0\\
		0&\cosh\left(\beta J \right)&-\sinh\left(\beta J \right)&  0\\0&-\sinh\left(\beta J \right)&\cosh\left(\beta J \right)&  0\\0&0&0&\exp\left(\beta B \right)\end{array}}\right).
\end{equation}
To determine the bounds on simultaneous estimation of parameters $B$ and $T$, we need to derive the formula
\begin{equation}
\left(\rho\otimes\openone+\openone\otimes\rho^{*} \right)^{-1}=\left( {\begin{array}{*{20}{c}}
		\varGamma_{11} &0_{4\times4}&0_{4\times4}&0_{4\times4}\\
		0_{4\times4}&\varGamma_{22}&\varGamma_{23}&  0_{4\times4}\\0_{4\times4}&\varGamma_{32}&\varGamma_{33}&  0_{4\times4}\\0_{4\times4}&0_{4\times4}&0_{4\times4}&\varGamma_{44}\end{array}}\right),
\end{equation}
with the quantities $\varGamma_{ij}$ ($i,j=1,2,3,4$) are
\begin{equation}
\varGamma_{11}=\left( {\begin{array}{*{20}{c}}
		\alpha &0&0&0\\
		0&\xi&\delta&  0\\0&\delta&\xi&  0\\0&0&0&\lambda\end{array}}\right),\hspace{2cm}\varGamma_{22}=\varGamma_{33}=\left({\begin{array}{*{20}{c}}
		\xi &0&0&0\\
		0&\kappa&\eta&0\\0&\eta&\kappa&  0\\0&0&0&\vartheta\end{array}}\right),
\end{equation}
and
\begin{equation}
	\varGamma_{44}=\left( {\begin{array}{*{20}{c}}
			\lambda &0&0&0\\
			0&\vartheta&\tau&0\\0&\tau&\vartheta&  0\\0&0&0&\alpha\end{array}}\right),\hspace{2cm}\varGamma_{23}=\varGamma_{32}=\left({\begin{array}{*{20}{c}}
			\mu&0&0&0\\
			0&\eta&\varpi&0\\0&\varpi&\eta&  0\\0&0&0&\tau\end{array}}\right),
\end{equation}
with the entries given by
\begin{equation}
	\alpha = {e^{\beta B}}\left( {\cosh\left( \beta B \right) + \cosh\left( \beta J \right)} \right),\hspace{1.5cm}\xi= 1 + {e^{\beta B}}\cosh\left( \beta J \right),
\end{equation}
\begin{equation}
	\delta = {e^{\beta B}}\sinh\left( \beta J \right), \hspace{1cm} \lambda = 1 + \frac{{\cosh\left( \beta J \right)}}{{\cosh\left( \beta B \right)}},\hspace{1cm} \tau = \left( {\cosh\left( \beta B \right) - \sinh\left( \beta B \right)} \right)\sinh\left( \beta J \right),
\end{equation}
\begin{equation}
	\kappa = \frac{1}{4}\left( {\cosh\left( \beta B \right) + \cosh\left( \beta J \right)} \right)\left( {3 + \cosh\left( 2\beta J \right)} \right){\mathop{\rm sech}\nolimits} \left( \beta J \right),
\end{equation}
\begin{equation}
	\eta = \frac{1}{2}\left( {\cosh\left( \beta B \right) + \cosh\left( \beta J \right)} \right){\mathop{\rm \sinh}\nolimits} \left( \beta J \right), \hspace{1.5cm}\vartheta = 1 + {e^{ - \beta B}}\cosh\left( \beta J \right),
\end{equation}
\begin{equation}
	\mu= {e^{\beta B}}\sinh\left( \beta J \right), \hspace{1.5cm}\varpi = \frac{1}{2}\left( {\cosh\left( \beta B \right) + \cosh\left( \beta J \right)} \right)\tanh\left( \beta J \right)\sinh\left( \beta J \right).
\end{equation}
In the estimated parameter space basis $\left\lbrace\left|B \right\rangle,\left|T\right\rangle\right\rbrace$, the quantum Fisher information matrix via the vectorization method is established as follows
\begin{eqnarray}
	\mathbb{F}_{\boldsymbol{\boldsymbol{ \theta}}}= \left[ {\begin{array}{*{20}{c}}
			{2{\rm vec}{{\left[ {{\partial _B}\rho } \right]}^T}^{}{\left(\rho\otimes\openone+\openone\otimes\rho^{*} \right)^{-1}}{\rm vec}\left[ {{\partial _B}\rho } \right]}&{2{\rm vec}{{\left[ {{\partial _B}\rho } \right]}^T}{\left(\rho\otimes\openone+\openone\otimes\rho^{*} \right)^{-1}}{\rm vec}\left[ {{\partial _T}\rho } \right]}\\
			{2{\rm vec}{{\left[ {{\partial _T}\rho } \right]}^T}{\left(\rho\otimes\openone+\openone\otimes\rho^{*} \right)^{-1}}{\rm vec}\left[ {{\partial _B}\rho } \right]}&{2{\rm vec}{{\left[ {{\partial _T}\rho } \right]}^T}{\left(\rho\otimes\openone+\openone\otimes\rho^{*} \right)^{-1}}{\rm vec}\left[ {{\partial _T}\rho } \right]}
	\end{array}} \right].
\end{eqnarray}
By a straightforward calculation, one can check that
\begin{eqnarray}
	{\rm vec}\left[ {{\partial _T}\rho } \right] = {\left( {{\partial _T}\chi,0,0,0,0,{\partial _T}\gamma,{\partial _T}\varsigma,0,0,{\partial _T}\varsigma,{\partial _T}\gamma,0,0,0,0,{\partial _T}\Pi} \right)^T},\label{113}
\end{eqnarray}
and
\begin{eqnarray}
	{\rm vec}\left[ {{\partial _B}\rho }\right] = {\left( {{\partial _B}\chi,0,0,0,0,{\partial _B}\gamma,{\partial _B}\varsigma,0,0,{\partial _B}\varsigma,{\partial _B}\gamma,0,0,0,0,{\partial _B}\Pi} \right)^T},\label{114}
\end{eqnarray}
where the quantities $\chi$, $\Pi$, $\zeta$ and $\gamma$ in the above equations are given by  
\begin{equation}
	\chi = \frac{{{e^{ - \beta B}}}}{{2\left( {\cosh\left( \beta B \right) + \cosh\left( \beta J \right)} \right)}}, \hspace{1.5cm}\Pi = \frac{{{e^{\beta B}}}}{{2\left( {\cosh\left( \beta B \right) + \cosh\left( \beta J \right)} \right)}},
\end{equation}
\begin{equation}
	\zeta = \frac{{ - \sinh\left( \beta J \right)}}{{2\left( {\cosh\left( \beta B \right) + \cosh\left( \beta J \right)} \right)}},\hspace{1.5cm}\gamma = \frac{{\cosh\left( \beta J \right)}}{{2\left( {\cosh\left( \beta B \right) + \cosh\left( \beta J \right)} \right)}}.
\end{equation}
Insertion of the expressions given in Eq.(\ref{113}) and Eq.(\ref{114}) into Eq.(9) yields the inverse of the quantum Fisher information matrix
\begin{equation}
	\mathbb{F}_{\boldsymbol{\boldsymbol{ \theta}}}^{ - 1} = \frac{1}{{\det \left( \mathbb{F}_{\boldsymbol{\boldsymbol{ \theta}}} \right)}}\left[ {\begin{array}{*{20}{c}}
			\boldsymbol{{\cal F}}_{TT}&{ - {\boldsymbol{{\cal F}}_{BT}}}\\
			{ - {\boldsymbol{{\cal F}}_{BT}}}&{{\boldsymbol{{\cal F}}_{BB}}}
	\end{array}} \right],
\end{equation}
where their entries coefficients are given by
\begin{equation}
	\boldsymbol{{\cal F}}_{TT} = \frac{{{{\rm{e}}^{ - 2\beta B}}\left( {1 + {{\rm{e}}^{2\beta B}} + 2{{\rm{e}}^{\beta B}}\cosh \left( \beta J \right)} \right)}}{{4{T^4}{{\left( {\cosh\left( \beta B \right) + \cosh\left( \beta J \right)} \right)}^3}}}\left(\begin{array}{l}
		\left( {1 + {{\rm{e}}^{2\beta B}}} \right)\left( {{B^2} + {J^2}} \right)\cosh\left( \beta J\right) +\\
		2\left( {{{\rm{e}}^{\beta B}}\left({{B^2} + {J^2}}\right)- B\left( { - 1 + {{\rm{e}}^{2\beta B}}} \right)J\sinh\left( \beta J \right)} \right)
	\end{array}\right),
\end{equation}
\begin{equation}
	\boldsymbol{{\cal F}}_{BB} = \frac{{2{{\rm{e}}^{\beta B}}\left( {2{{\rm{e}}^{\beta B}} + \left( {1 + {{\rm{e}}^{2\beta B}}} \right)\cosh\left( \beta J\right) } \right)}}{{{T^2}{{\left( {1 + {{\rm{e}}^{2\beta B}} + 2{{\rm{e}}^{\beta B}}\cosh\left( \beta J\right) } \right)}^2}}},
\end{equation}
and
\begin{equation}
	\boldsymbol{{\cal F}}_{BT}= {F_{TB}} = \frac{{2{{\rm{e}}^{\beta B}}\left( {2B{{\rm{e}}^{\beta B}} + B\left( {1 + {{\rm{e}}^{2\beta B}}} \right)\cosh\left(  \beta J \right)  - \left( { - 1 + {{\rm{e}}^{2\beta B}}} \right)J\sinh \left(  \beta J \right) } \right)}}{{{T^3}{{\left( {1 + {{\rm{e}}^{2\beta B}} + 2{{\rm{e}}^{\beta B}}\cosh\left( \beta J \right) } \right)}^2}}}.
\end{equation}
In any given quantum multiparameter metrology strategy, the quantum Cramer-Rao bound provided us with the lower limit of the covariance matrix of estimators $\hat{\boldsymbol{ \theta}}=\left(B,T \right)$. This limit reads as follows 
\begin{equation}
{\rm Cov}\left(\hat{\boldsymbol{ \theta}}\right)\equiv\left[ {\begin{array}{*{20}{c}}
		{\rm Var}\left( B\right)&{\rm Cov}\left( B,T\right)\\
		{\rm Cov}\left( T,B\right)&{\rm Var}\left( T\right)
\end{array}} \right]\geq 	\mathbb{F}_{\boldsymbol{\boldsymbol{ \theta}}}^{ - 1}. 
\end{equation}
By comparing the coefficients on both sides of the quantum Cramer-Rao inequality, we then get
\begin{eqnarray}
	{\rm Var}\left( B\right) \ge \frac{{{\boldsymbol{{\cal F}}_{TT}}}}{{\det \left( \mathbb{F}_{\boldsymbol{\boldsymbol{ \theta}}} \right)}}, \hspace{2cm}{\rm and}\hspace{2cm}{\rm Var}\left( T\right) \ge \frac{{{\boldsymbol{{\cal F}}_{BB}}}}{{\det \left( \mathbb{F}_{\boldsymbol{\boldsymbol{ \theta}}} \right)}}.
\end{eqnarray}
As previously shown, the saturation of these above inequalities provides us the highest accuracy on the estimation of the parameters $B$ and $T$. Therefore, the optimal analytical expressions of the precision are coincident with the minimum variances of the estimated parameters. Under these conditions, ${\mathop{\rm Var}}\left(B\right)_{\min }$ for the external magnetic field parameter and ${\mathop{\rm Var}}\left(T\right)_{\min }$ for the temperature parameter yield the following results
\begin{equation}
	{\mathop{\rm Var}}\left(T\right)_{\min } = \frac{{{{\rm{e}}^{ - \beta B}}{T^4}\left( {2{{\rm{e}}^{\beta B}} + \left( {1 + {{\rm{e}}^{2\beta B}}} \right)\cosh\left( \beta J\right)} \right)}}{{2{J^2}}}. \label{varminT2}
\end{equation}
and
\begin{equation}
	{\rm{Var}}{\left(B\right)_{\min }} = \frac{{{{\rm{e}}^{ - 4\beta B}}{T^2}{{\left( 1 + {{\rm{e}}^{2\beta B}} + 2{{\rm{e}}^{\beta B}}\cosh \left( \beta J \right) \right)}^3}}}{{16{J^2}{{\left( {\cosh \left( \beta B \right) + \cosh\left(\beta J\right)} \right)}^3}}}\left( \begin{array}{l}
		\left( {1 + {{\rm{e}}^{2\beta B}}} \right)\left( {{B^2} + {J^2}} \right)\cosh\left( \beta J\right) + \\
		2\left( {{{\rm{e}}^{\beta B}}\left( {{B^2} + {J^2}} \right) - B\left( { - 1 + {{\rm{e}}^{2\beta B}}} \right)J\sinh\left(\beta J \right)} \right)
	\end{array} \right). \label{varminB}
\end{equation}
Straightforward calculations leads to find also the analytical expressions of the symmetric logarithmic derivatives corresponding respectively to the parameters $T$ and $B$ as
\begin{equation}
	{L_T} = 2\left[ {\begin{array}{*{20}{c}}
			{\alpha\,{\partial _T}\,\chi}&0&0&0\\
			0&{(\kappa + \varpi){\mkern 1mu} {\partial _T}\,\gamma + 2\eta{\mkern 1mu} {\partial _T}\zeta}&{(\kappa + \varpi){\mkern 1mu} {\partial _T}\varsigma + 2\eta{\mkern 1mu} {\partial _T}\gamma}&0\\
			0&{(\kappa + \varpi){\mkern 1mu} {\partial _T}\varsigma + 2\eta{\mkern 1mu} {\partial _T}\gamma}&{(\kappa + \varpi){\mkern 1mu} {\partial _T}\gamma + 2\eta{\mkern 1mu} {\partial _T}\varsigma}&0\\
			0&0&0&{\alpha{\mkern 1mu} {\partial _T}\Pi}
	\end{array}} \right],
\end{equation}
and
\begin{equation}
	{L_B} = 2\left[ {\begin{array}{*{20}{c}}
			{\alpha\,{\partial _B}\,\chi}&0&0&0\\
			0&{(\kappa + \varpi){\mkern 1mu} {\partial _B}\,\gamma + 2\eta{\mkern 1mu} {\partial _B}\varsigma}&{(\kappa + \varpi){\mkern 1mu} {\partial _B}\varsigma + 2\eta{\mkern 1mu} {\partial _B}\gamma}&0\\
			0&{(\kappa + \varpi){\mkern 1mu} {\partial _B}\varsigma + 2\eta{\mkern 1mu} {\partial _B}\gamma}&{(\kappa + \varpi){\mkern 1mu} {\partial _B}\gamma + 2\eta{\mkern 1mu} {\partial _B}\varsigma}&0\\
			0&0&0&{\alpha{\mkern 1mu} {\partial _B}\Pi}
	\end{array}} \right].
\end{equation}
Generally, the multi-parameter estimation protocol is harder than the single-parameter estimation protocol since we cannot always achieve the optimal estimation of multiple parameters simultaneously encrypted in a given quantum system. This is mainly due to the incompatibility of quantum measurements \cite{Razavian2020}. Consequently, the quantum limits of accuracy cannot usually be achieved. To saturate the quantum Cramér-Rao bound in multiparameter estimation problems, we need to know the influence of this incompatibility on our estimation problem. This is efficiently done by using a recently introduced quantity named quantumness \cite{Carollo2019,Carollo2018}. It is a scalar quantity that evaluates the asymptotic incompatibility of a simultaneous estimate of several parameters and is expressed as follows
\begin{equation}
{\cal R}_{\boldsymbol{ \theta}}:=\parallel2i\mathbb{F}_{\boldsymbol{\boldsymbol{ \theta}}}^{-1}U_{\boldsymbol{ \theta}}\parallel_{\infty},\label{RU}
\end{equation}
where $\parallel Z\parallel_{\infty}$ stands for the largest eigenvalue of a matrix $Z$ and $U_{\boldsymbol{ \theta}}$ is the mean Uhlmann curvature matrix defined by
\begin{equation}
U_{\theta_{\mu}\theta_{\nu}}=-\frac{i}{4}{\rm Tr}\left(\rho_{\theta}\left[\hat{L}_{\theta_{\mu}},\hat{L}_{\theta_{\nu}}\right]\right).
\end{equation}
The values of the quantumness ${\cal R}_{\boldsymbol{\theta}}$ are included in the range $0\leq{\cal R}_{\boldsymbol{\theta}}\leq1$. The Cramer-Rao inequality (\ref{BCR}) is not saturated when the incompatibility between the measurements of the simultaneously estimated parameters is maximal, which corresponds to the quantumness maximal with ${\cal R}_{\boldsymbol{\theta}}=1$. Contrary, the condition ${\cal R}_{\boldsymbol{\theta}}=0$ ensures the saturation of Cramér-Rao bound and the parameterization model is asymptotically compatible. Furthermore, we get
\begin{equation}
U_{\theta_{\mu}\theta_{\nu}}=0\hspace{0.5cm}\Leftrightarrow\hspace{0.5cm}{\cal R}_{\boldsymbol{ \theta}}=0,
\end{equation}
and in the special case of a two parameter simultaneous estimation, equation (1\ref{RU}) becomes
\begin{equation}
{\cal R}_{\boldsymbol{ \theta}}=\sqrt{\frac{{\rm Det\left(2U_{\theta_{\mu}\theta_{\nu}}\right) }}{{\rm Det\left(\mathbb{F}_{\boldsymbol{\boldsymbol{ \theta}}}\left(\theta_{\mu},\theta_{\nu} \right)\right) }}}.
\end{equation}
\section{Interferometric phase estimation and non-classical correlations}
Very recently, a major effort has been undertaken to evaluate the dynamics of quantum Fisher information to establish the validity of quantum entanglement in various quantum metrology scenarios and models. In unitary processes, it has been shown that entanglement leads to a significant improvement in the accuracy of parameter estimation. In fact, the quantum entanglement can be exploited as a quantum resource to exceed the standard quantum limit and achieve the Heisenberg limit \cite{Giovannetti2004}. Furthermore, the gain of information that becomes noise-resistant is essential to promising entanglement in quantum estimation theory \cite{Sen2016,Petz2011}. This fact raises an important question; Can quantum correlations beyond entanglement be related to the estimation precision in quantum metrology protocols? Is it possible to quantify the quantum correlations in terms of quantum Fisher information? Several studies have recently been conducted in this direction. Indeed, a quantum correlation quantifier called the local quantum Fisher information has been recently introduced to understand the role of quantum correlation beyond entanglement in a black-box quantum metrology task \cite{Kim2018}. This quantifier is related directly to the quantum Fisher information and also provides the accuracy of an interferometric phase estimate. Local quantum uncertainty \cite{Girolami2013}, as another quantum correlation measure, can also be related to the precision in phase estimation protocols. This supports the idea that quantum Fisher information is adding a new tool to the existing list of quantum correlation quantifiers in bipartite quantum systems.
\begin{figure}[h]
	\includegraphics[scale=0.3]{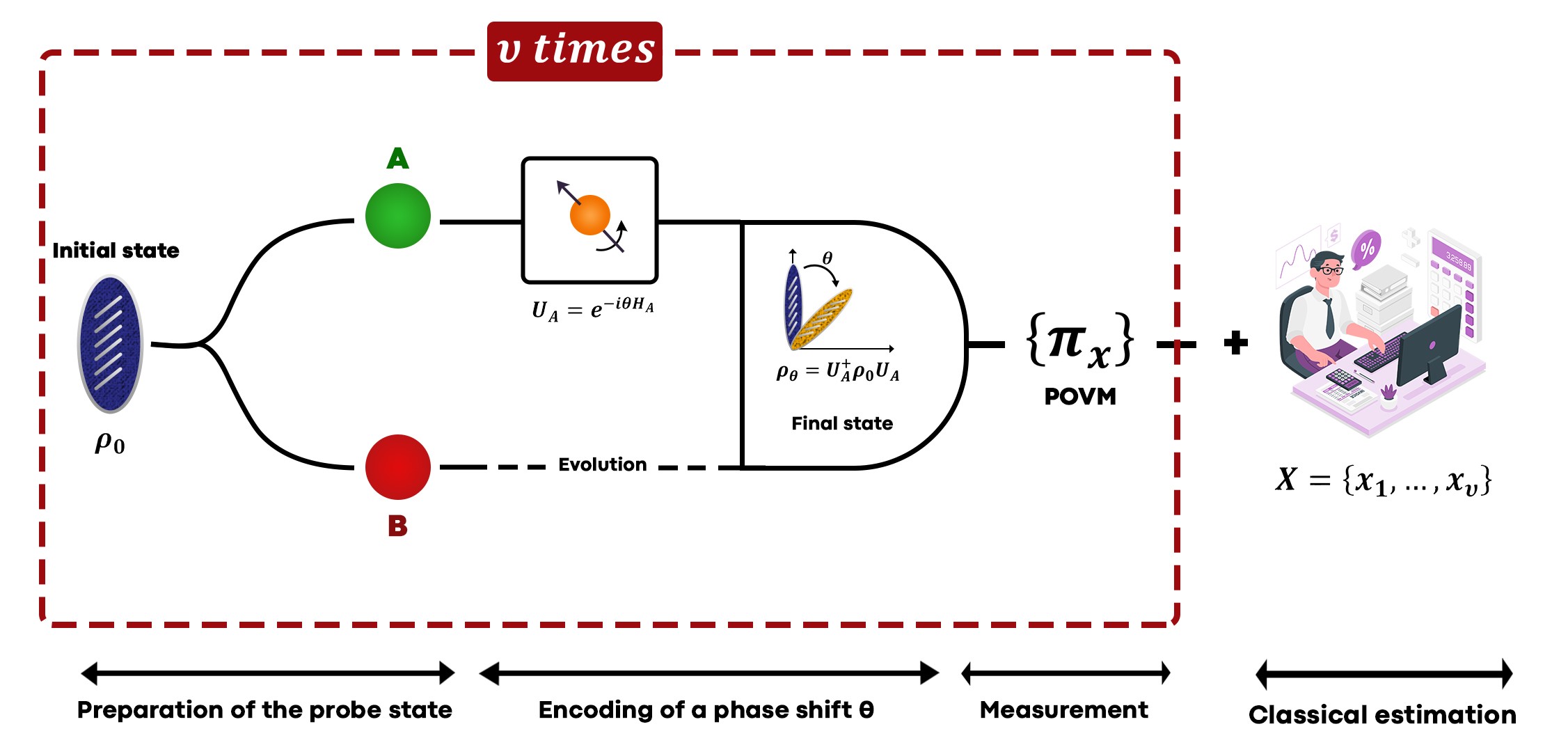}
	\caption{Conceptual scheme of a complete quantum phase estimation metrological process. The various parameters are defined in the main text.}\label{Fig2}
\end{figure}

A complete quantum metrology process generally includes four steps (Fig.\ref{Fig2}); First, the preparation of the probe state. Second, the estimated parameter parameterization via unitary evolution. Third, a measurement on the final state via POVM measurements and fourth step consists of data analysis via classical estimation. This last step has been well studied in classical statistics, therefore, the main concern of quantum estimation is the first three steps. Here, we focus on the first step in which we examine how non-classical correlations in the input state improve the accuracy of the estimated parameters in the interferometric phase estimation protocol.
 \subsection{Local quantum Fisher information as a measure of non-classical correlations}
A major challenge for information theorists is to find a new measure of quantum correlations via quantum Fisher information. Actually, local quantum Fisher information has been introduced to treat discord-type quantum correlations in interferometric phase estimation \cite{Kim2018}. This quantifier allows us to better understand how non-classical correlations contribute to establishing accuracy in quantum estimation theory. It meets the criteria identifying discord type quantifiers that any good quantum correlation quantifier should satisfy. Indeed, it is non-negative and vanishes only for the classically correlated states. Also, it is invariant under any local unitary operation and coincides with the geometric quantum discord for any pure quantum states.\par

Consider first a bipartite quantum state $\rho^{AB}$ of dimension $d_{A}\times 
d_{B}$ acting on a Hilbert space $H_{A}\otimes H_{B}$, and let $K_{A}=K_{A}\otimes\openone_{B}$ be a local observable on a single part $A$. The dynamics of the density matrix $\rho^{AB}$ is governed by the local phase shift transformation $e^{-i\theta K_{A}}$, i.e.,  $\rho_{\theta}^{AB}=e^{-i\theta K_{A}}\rho^{AB}e^{i\theta K_{A}}$. Quantum Fisher information via local measurements and via local observables, ${\cal F}\left(\rho^{AB},K_{A} \right)$, is called local quantum Fisher information on the sub-system $A$. When ${\cal F}\left(\rho^{AB},K_{A} \right)=0$, then no information can be obtained by measurement because no change in the evolution will be observed. Building on this, the local quantum Fisher information captures the amount of quantum correlations that exist in a given quantum state and allows to quantify the non-classical correlations in terms of quantum Fisher information \cite{Kim2018}. It is defined as the minimum quantum Fisher information over all local observables with respect to the subsystem $A$ and given by 
\begin{equation}
	{\cal Q}_{F}\left(\rho^{AB}\right)=\min_{K_{A}}{\cal F}\left(\rho^{AB},K_{A}\right).\label{LQFI}
\end{equation}
The advantage of using this measure compared to other quantum discord quantifiers relies on the possibility to obtain the closed analytical form. For $2\times d$ quantum systems with ${\rm dim}\left({\cal H}_{A}\right)=2$ and ${\rm dim}\left({\cal H}_{B}\right)=d$, it is simple to show that the local quantum Fisher information (\ref{LQFI}) reduces to
\begin{equation}
	{\cal Q}_{F}\left(\rho^{AB}\right)=\min_{K_{A}}\left[{\rm Tr}\left(\rho K_{A}^{2} \right)-\sum_{i\neq j}\frac{2p_{i}p_{j}}{p_{i}+p_{j}}|\left\langle\psi_{i}\right| K_{A}\left|\psi_{j} \right\rangle|^{2}\right],\label{497}
\end{equation}
where we used the spectral decomposition of the density matrix $\rho^{AB}$, i.e., $\rho^{AB}=\sum_{i}p_{i}\left|\psi_{i} \right\rangle\left\langle\psi_{i}\right|$ with $p_{i}\geq0$ and $\sum_{i}p_{i}=1$.\par

Basically, any qubit can be represented geometrically in the Bloch sphere and the general form of a local Hamiltonian can be described as $K_{A}=\vec{\sigma}\vec{r}$, with $|\vec{r}|=1$ and $\vec{\sigma}=\left(\sigma_{x},\sigma_{y},\sigma_{z}\right)$ are the usual Pauli matrices (the generators of the $su(2)$ algebra). Therefore, we can show that ${\rm Tr}\left(\rho K_{A}^{2} \right)=1$ and the second part of the equation (\ref{497}) is expressed as below
\begin{align}                
	\sum_{i\neq j}\frac{2p_{i}p_{j}}{p_{i}+p_{j}}|\left\langle\psi_{i}\right| K_{A}\left|\psi_{j} \right\rangle|^{2}&=\sum_{i\neq j}\sum_{l,k=1}^{3}\frac{2p_{i}p_{j}}{p_{i}+p_{j}}\left\langle\psi_{i}\right| \sigma_{l}\otimes\openone_{B}\left|\psi_{j} \right\rangle\left\langle\psi_{j}\right| \sigma_{k}\otimes\openone_{B}\left|\psi_{i} \right\rangle\notag\\&=\vec{r}^{\dagger}.M.\vec{r},
\end{align}
where the components of the $3\times3$ symmetric matrix $M$ are determined by
\begin{equation}
	M_{lk}=\sum_{i\neq j}\frac{2p_{i}p_{j}}{p_{i}+p_{j}}\left\langle\psi_{i}\right| \sigma_{l}\otimes\openone_{B}\left|\psi_{j} \right\rangle\left\langle\psi_{j}\right| \sigma_{k}\otimes\openone_{B}\left|\psi_{i} \right\rangle. \label{MatrixM}
\end{equation}
The minimization of ${\cal F}\left(\rho^{AB},K_{A}\right)$ requires maximizing the quantity $\vec{r}^{\dagger}.M.\vec{r}$ on all unit vectors $\vec{r}$. Here, its maximum value corresponds to the maximum eigenvalue of $M$. Therefore, the analytical expression of the local quantum Fisher information ${\cal Q}_{F}\left(\rho^{AB}\right)$, which quantifying the pairwise quantum correlation in the given quantum bipartite state, yields as
\begin{equation}
	{\cal Q}_{F}\left(\rho^{AB}\right)=1-\lambda_{\rm max}\left(M \right),
\end{equation}
with $\lambda_{\rm max}$ denotes the maximal eigenvalue of the symmetric matrix $M$ whose entries are given in equation (\ref{MatrixM})
\subsection{Local quantum uncertainty as a discord-like quantity}
The physicists study nature by making measurements and predicting their results. In a classical world, the error bands are exclusively due to technological limitations, so it is possible to measure any two observables with arbitrary precision. However, such a type of measurement is not always possible in quantum systems since quantum mechanics stipulates that two non-commutable observables cannot be measured jointly with arbitrary precision even if one could access a flawless measuring device. Moreover, the measurement of a quantum system is completely different from that of a classical system, because not only the result obtained is probabilistic, but also the state of the quantum system is modified during this process, and the objective is to determine, in a more precise way, the result of this measurement. The idea of accuracy is directly related to the uncertainty associated with the result of this measurement and the uncertainty relation gives the statistical nature of the errors in this kind of measurement.\par
In standard practice and in the framework of the quantum mechanical formalism, the uncertainty of an observable $K$ in a quantum state $\rho$ is generally quantified by the variance which is given by the following relation
\begin{equation}
	{\rm Var}\left(\rho,K\right):={\rm Tr}\left(\rho K_{0}^{2} \right)= {\rm Tr}\left( \rho K^{2}\right)-{\rm Tr}\left(\rho K\right)^{2},\label{Vr}
\end{equation}
where ${\rm Tr}$ denotes the trace and $K_{0}=K-{\rm Tr}\rho K$. For the pure states $\rho=\left|\psi \right\rangle\left\langle\psi \right|$, the variance reduces to
\begin{equation}
	{\rm Var}\left(\left|\psi \right\rangle\left\langle\psi \right|,K\right)=\left\langle\psi \right| K^{2}\left| \psi\right\rangle -\left\langle\psi \right| K\left| \psi\right\rangle^{2}.
\end{equation}
Indeed, the quantification of uncertainty in terms of variance is well adapted when the states are pure. However, this uncertainty includes classical ignorance and a part of quantum nature for mixed states. The mixture of states is responsible for the classical part and the quantum part resulting from the non-commutativity between the studied state and the measured observable. Thus, the variance can have contributions from both quantum and classical sources, i.e., ${\rm Var}\left(\rho\right)\equiv{\rm Var}_{q}\left(\rho\right)+{\rm Var}_{c}\left(\rho\right)$. There are still several methods to quantify the uncertainty, such as those based on entropy measurements which are widely used as indicators of uncertainty in quantum information theory. This type of measurement is still insufficient and limited because these quantifiers are always influenced by the mixed state of the studied state. In order to determine the quantum part of the variance (\ref{Vr}), Wigner and Yanase introduced the notion of skew information as \cite{Wigner1963,Luo2003}
\begin{equation}
	{\cal I}\left( {\rho ,K} \right): = {{\cal I}_\rho }\left( K \right): = \frac{1}{2}{\rm Tr}\left\{ {{{\left( {i\left[ {\sqrt \rho  ,K} \right]} \right)}^2}} \right\},\label{IS}
\end{equation}
where $\left[.,.\right]$ stands for the commutator. Importantly, unlike the variance, the Wigner-Yanase skew information is not affected by classical mixing. The latter quantifies the degree of non-commutativity between a quantum state $\rho$ and an observable $K$ that can be considered as a Hamiltonian or any other conserved quantity. Furthermore, the skew information provides an alternative measure of the information content of the studied state with respect to non-commutative observable with the conserved quantity $K$. In other words, it represents the amount of information contained in the quantum state $\rho$ that is not accessible by the measurement of observable $K$. The equation (\ref{IS}) can be simplified as 
\begin{align}
	{\cal I}\left( {\rho ,K} \right)& =-\frac{1}{2}{\rm Tr}\{\left(\sqrt{\rho}K-K\sqrt{\rho}\right)^{2}\}\notag \\&=- \frac{1}{2}\left\{ {{\rm Tr}\left( {\sqrt \rho  K\sqrt \rho  K} \right) - {\rm Tr}\left( {K\rho K} \right) - {\rm Tr}\left( {\sqrt \rho  {K^2}\sqrt \rho  } \right) + {\rm Tr}\left( {K\sqrt \rho  K\sqrt \rho  } \right)} \right\} \notag \\
	& = {\rm Tr}\left( {\rho {K^2}} \right) - {\rm Tr}\left( {\sqrt \rho  K\sqrt \rho  K} \right). \label{SII} 
\end{align}
Very recently, Girolami and co-workers defined the local quantum uncertainty as the minimum skew information that can be obtained by a single local measurement \cite{Girolami2013}. More precisely, for a bipartite quantum state $\rho_{AB}$, we consider the local observable $K^{\Lambda}=K_{A}^{\Lambda}\otimes\openone_{B}$ such that $K_{A}^{\Lambda}$ is a Hermitian operator acting on states of the $A$ subsystem and having a non-degenerate spectrum $\Lambda$. The local quantum uncertainty with respect to the subsystem $A$ is given by
\begin{equation}
	{\cal U}_{A}^{\Lambda}\left( \rho_{AB}\right):=\min_{K_{A}^{\Lambda}}{\cal I}\left(\rho_{AB}, K_{A}^{\Lambda}\otimes\openone_{B}\right), \label{DLQU}
\end{equation}
where the minimization involves all the local observables that act on the subsystem $A$. Indeed, $\Lambda$ here is non-degenerate because it corresponds to the maximal informational observables of subsystem $A$. This quantifier has strong reasons to be considered as a faithful measure of the quantum correlations of bipartite quantum states. Indeed, the local quantum uncertainty is invariant under local unitary transformations, is not increasing under local operations on the subsystem $B$, vanish if and only if the quantum state is classically correlated \cite{Girolami2013}. For pure states, the local quantum uncertainty is a monotonic entanglement. As a result, the local quantum uncertainty satisfies all the physical requirements for a quantum correlation measure.\par
To evaluate the minimum in equation (\ref{DLQU}), there is an important requirement for the optimization observables that minimize the skew information, namely that they must have a fixed non-degenerate spectrum. In practice, the observables with this spectrum acting on the subsystem $A$ can be parameterized by
\begin{equation}
	K_{A}^{\Lambda}=V_{A}\Lambda V_{A}^{\dagger},
\end{equation}
where $V_{A}$ defines the measurement basis that can be arbitrarily modified on the special unit group of the subsystem $A$. If we restrict our discussion to the case of bipartite quantum systems of qubit-qudit type where the subsystem $A$ has dimension $2$ and the subsystem $B$ has dimension $d$, the non-degenerate fixed-spectrum qubit observable $\sigma_{3}$ can be parametrized by $K_{A}=V_{A}\sigma_{3} V_{A}^{\dagger}=\vec{s}.\vec{\sigma}$, where $\Arrowvert\vec{s}\Arrowvert=1$ and $\sigma_{i}$ are the Pauli matrices, also representing the generators of the $su\left(2\right)$ algebra. 
By selecting this spectrum, the local quantum uncertainty is normalized to unity for maximally entangled pure states. Then, the $\Lambda$ spectrum can be removed in the equation (\ref{DLQU}) and the local quantum uncertainty is equivalent to
\begin{align}
	{\cal U}_{A}\left(\rho_{AB}\right)&=\min_{\vec{s}}{\cal I}\left(\rho_{AB},\vec{s}.\vec{\sigma}\otimes\openone_{d} \right)\notag\\&=\min_{\vec{s}}\sum_{ij}s_{i}s_{j}\left[{\rm Tr}\{\rho_{AB}\sigma_{i}\sigma_{j}-\sqrt{\rho_{AB}}\left(\sigma_{i}\otimes\openone_{d} \right)\sqrt{\rho_{AB}}\left(\sigma_{j}\otimes\openone_{d}\right)\} \right]\notag\\&=1-\max\sum_{ij}s_{i}s_{j}\left[{\rm Tr}\{\sqrt{\rho_{AB}}\left(\sigma_{i}\otimes\openone_{d} \right)\sqrt{\rho_{AB}}\left(\sigma_{j}\otimes\openone_{d}\right)\} \right]\notag\\&=1-\max\left[ \vec{s}^{\dagger}{\cal W}\vec{s}\right],\label{LQUANLY}
\end{align}
where ${\cal W}$ is a $3\times3$ symmetric matrix whose elements are
\begin{equation}
	w_{ij}={\rm Tr}\{\sqrt{\rho_{AB}}\left(\sigma_{i}\otimes\openone_{B} \right)\sqrt{\rho_{AB}}\left(\sigma_{j}\otimes\openone_{B}\right)\},\label{matriceW}
\end{equation}
with $i,j=1,2,3$. 
As we have already seen in the minimization procedure of the local quantum Fisher information, it is necessary to maximize the quantity ${\cal W}$ on all unit vectors $\vec{s}$ to minimize the skew information. Its value coincides with the maximal eigenvalue of the matrix ${\cal W}$. Consequently, a compact formula for the local quantum uncertainty is given by
\begin{equation}
	{\cal U}_{A}\left(\rho_{AB}\right)=1-\max\{\xi_{1},\xi_{2},\xi_{3}\},\label{lqusimple}
\end{equation}
where $\xi_{i}$ represent the eigenvalues of the matrix ${\cal W}=\left(w_{ij} \right)_{3\times3}$. Hence, it is clear that the calculation of the local quantum uncertainty requires the calculation of the elements $w_{ij}$ of the matrix ${\cal W}$. Thus, this measure has the advantage of being simple when it comes to calculating the maximum on the parameters related to the measurements as it must be done to calculate the quantum discord. Moreover, the local quantum uncertainty is reduced to the monotonic entanglement for pure states. In fact, it reduces to the linear entropy of the measured reduced subsystem (i.e., concurrence). This interesting coincidence occurs only for quantum systems of dimension $2\otimes d$. However, for multipartite quantum embedded systems (for a large Hilbert space dimension), it is difficult to obtain a compact form of the local quantum uncertainty without complex optimization. To clarify this issue, we consider a pure state $\left| \psi \right\rangle=\sum_{mn}C_{mn}\left|mn \right\rangle $, the elements of the matrix $\cal W$ become 
\begin{equation}
	w_{ij}=\left\langle \psi\right|\sigma_{i}\otimes\openone_{d}\left|\psi\right\rangle\left\langle \psi\right|\sigma_{j}\otimes\openone_{d}\left|\psi\right\rangle.
\end{equation}
The first term of the above equation becomes as follows
\begin{align}
	\left\langle \psi\right|\sigma_{i}\otimes\openone_{d}\left|\psi\right\rangle=\sum_{mk}\left(C{C}^{\dagger} \right)_{km} \left\langle m\right|\sigma_i\left| k\right\rangle\label{yy}.
\end{align}
The reduced density matrix of subsystem $A$ can be rewritten in terms of the coefficients $C_{mn}$ as
\begin{align}
	\rho_{A}=\sum_{mkl} C_{ml}C_{lk}^{\dagger}\left| m\right\rangle\left\langle k\right|=\sum_{mkl}\left(CC^{\dagger} \right)_{mk}\left|m\right\rangle\left\langle k\right|. \label{rhoaa}
\end{align}
We then obtain
\begin{equation}
	\left\langle m\right|\rho_{A}\left| k\right\rangle =\left(CC^{\dagger} \right)_{mk}.\label{rr}
\end{equation}
We return to the expression (\ref{yy}) and using the equation (\ref{rr}), we can easily see that it reduces to the average value of $\sigma_{i}$ with respect to the base of the subsystem $A$. So we have

\begin{align}
	\left\langle\psi\right|\sigma_{i}\otimes\openone_{d}\left|\psi\right\rangle=\sum_{k}\left\langle k\right|\rho_{A}\sigma_{i}\left| k\right\rangle={\rm Tr}_{A}\left(\rho_{A}\sigma_{i} \right)=\left\langle \sigma_{i}\right\rangle_{A}.
\end{align}
This leads to $w_{ij}=\left\langle \sigma_{i}\right\rangle_{A}\left\langle \sigma_{j}\right\rangle_{A}$, and the matrix ${\cal W}$ can be written as
\begin{equation}
	{\cal W}=\left( {\begin{array}{*{20}{c}}
			{\left\langle \sigma_{1}\right\rangle_{A}^{2}}&{\left\langle \sigma_{1}\right\rangle_{A}\left\langle \sigma_{2}\right\rangle_{A}}&{\left\langle \sigma_{1}\right\rangle_{A}\left\langle \sigma_{3}\right\rangle_{A}}\\
			{\left\langle \sigma_{2}\right\rangle_{A}\left\langle \sigma_{1}\right\rangle_{A}}&{\left\langle \sigma_{2}\right\rangle_{A}^{2}}&{\left\langle \sigma_{2}\right\rangle_{A}\left\langle \sigma_{3}\right\rangle_{A}}\\
			{\left\langle \sigma_{3}\right\rangle_{A}\left\langle \sigma_{1}\right\rangle_{A}}&{\left\langle \sigma_{2}\right\rangle_{A}\left\langle \sigma_{3}\right\rangle_{A}}&{\left\langle \sigma_{3}\right\rangle_{A}^{2}}
	\end{array}} \right).
\end{equation}
By diagonalizing the matrix ${\cal W}$, we find a single non-zero eigenvalue which is written
\begin{equation}
	\xi_{\max}=\left\langle \sigma_{1}\right\rangle_{A}^{2}+\left\langle \sigma_{2}\right\rangle_{A}^{2}+\left\langle \sigma_{3}\right\rangle_{A}^{2},
\end{equation}
and the local quantum uncertainty for pure states becomes
\begin{equation}
	{\cal U}_{A}\left( \left|\psi \right\rangle \left\langle \psi\right| \right)=1-\left[\left\langle \sigma_{1}\right\rangle_{A}^{2}+\left\langle \sigma_{2}\right\rangle_{A}^{2}+\left\langle \sigma_{3}\right\rangle_{A}^{2} \right]. 
\end{equation}
Besides, the Fano-Bloch representation of the matrix (\ref{rhoaa}) is written $\rho_{A}=\frac{1}{2}\left(1+\vec{r}\vec{\sigma_{i}} \right)$, where $\vec{r}$ denotes the Bloch vector with
\begin{equation}
	\left\langle \sigma_{i}\right\rangle_{A}={\rm Tr}\left(\rho_{A}\sigma_{i} \right)=\frac{1}{2}\left({\rm Tr}\sigma_{i}+r_{j}{\rm Tr}\left(\sigma_{i}\sigma_{j} \right)\right)=r_{i}.  
\end{equation}
Similarly, we also have
\begin{equation}
	{\rm Tr}\left(\rho_{A}^{2} \right)=\frac{1}{2}\left(1+r_{i}^{2} \right)= \frac{1}{2}\left(1+\xi_{\max} \right).
\end{equation}
Therefore, for pure states $\rho_{AB}=\left|\psi \right\rangle \left\langle \psi\right|$, the local quantum uncertainty coincides with the linear entropy of entanglement. We have then 
\begin{equation}
	{\cal U}_{A}\left( \left|\psi \right\rangle \left\langle \psi\right| \right)=2\left[1-{\rm Tr} \left(\rho_{A}^{2} \right) \right]=S_{2}\left(\rho_{A} \right).
\end{equation}
Interestingly, the analytical expression of the local quantum uncertainty (\ref{LQUANLY}) can be applied to multi-qubit systems without any major difficulty. Indeed, we can always consider the multi-qubit system (with $N$ qubits) as a system of dimension $2\otimes d$ where $d=2\otimes...\otimes2$ can represent the remaining $N-1$ qubits as a quantum system with a Hilbert space of dimension $d$. Since there are several bipartitions for a multi-qubit system, it is possible that some bipartitions are quantum correlated, some classically correlated and some may be completely uncorrelated with other qubits. Therefore, we need to define the local quantum uncertainty for each bipartition. For each case, we obtain a symmetric matrix ${\cal W}_{k}$, so in total we obtain $N$ symmetric matrices. After calculating all these local quantum uncertainties, we can calculate the average local quantum uncertainty for a given multi-qubit state. For this purpose, we consider an arbitrary density matrix $\rho_{N}$ associated to a system describing $N$ qubits and we apply local measurements on each qubit. Then, the matrix elements of each bipartition $\rho_{k}$ (with $k=1,2,...,N$) are given by the following expressions
\begin{align}
	&{\hat{w}}_{ij}^{1}={\rm Tr}\{\sqrt{\rho_{N}}\left(\sigma_{i}\otimes\openone_{2}...\otimes\openone_{2} \right)\sqrt{\rho_{N}}\left(\sigma_{j}\otimes\openone_{2}...\otimes\openone_{2}\right)\},\notag\\&{\hat{w}}_{ij}^{2}={\rm Tr}\{\sqrt{\rho_{N}}\left(\openone_{2}\otimes\sigma_{i}...\otimes\openone_{2} \right)\sqrt{\rho_{N}}\left(\openone_{2}\otimes\sigma_{j}...\otimes\openone_{2}\right)\},\notag\\&\hspace{0.2cm}:\hspace{1cm}:\hspace{1cm}:\hspace{1cm}:\hspace{1cm}:\hspace{1cm}:\hspace{1cm}:\hspace{1cm}:\notag\\&\hspace{0.2cm}:\hspace{1cm}:\hspace{1cm}:\hspace{1cm}:\hspace{1cm}:\hspace{1cm}:\hspace{1cm}:\hspace{1cm}:\notag\\&{\hat{w}}_{ij}^{N}={\rm Tr}\{\sqrt{\rho_{N}}\left(\openone_{2}\otimes\openone_{2}...\otimes\sigma_{i} \right)\sqrt{\rho_{N}}\left(\openone_{2}\otimes\openone_{2}...\otimes\sigma_{j}\right)\}.
\end{align}
Then, the local quantum uncertainties associated with each bipartition are determined as follows
\begin{align}
	&{\cal U}_{1/23...N}\left( \rho_{N}\right)=1-\max\{\xi_{1}^{1},\xi_{2}^{1},\xi_{3}^{1}\},\notag\\&{\cal U}_{2/13...N}\left( \rho_{N}\right)=1-\max\{\xi_{1}^{2},\xi_{2}^{2},\xi_{3}^{2}\},\notag\\&\hspace{0.2cm}:\hspace{1cm}:\hspace{1cm}:\hspace{1cm}:\hspace{1cm}:\hspace{1cm}:\notag\\&\hspace{0.2cm}:\hspace{1cm}:\hspace{1cm}:\hspace{1cm}:\hspace{1cm}:\hspace{1cm}:\notag\\&{\cal U}_{N/12...N-1}\left( \rho_{N}\right)=1-\max\{\xi_{1}^{N},\xi_{2}^{N},\xi_{3}^{N}\},
\end{align}
where $\xi_{i}^{k}$ (with $i=1,2,3$) are the eigenvalues of the symmetric matrices $\left({\cal W}_{k}\right)_{3\times3}$ which can be determined easily. The local quantum uncertainty, which provides us the quantity of the global quantum correlations in the global state $\rho_{N}$, can be viewed as the average value of the local quantum uncertainties of each bipartition, i.e.,
\begin{equation}
	{\cal U}\left( \rho_{N}\right)=\frac{1}{N}\left(\sum_{k=A}^{N}{\cal U}_{k/N_{k}} \right)\equiv\frac{1}{N}\left(\sum_{k=A}^{N}{\cal U}_{k/12...\bar{k}...N} \right),
\end{equation}
with $N_{k}=12...\bar{k}...N$ are the remaining ($N-1$)-qubits excluding the $k$-qubit.\par

Admittedly, the quantum correlation quantification in qudit-qudit systems ($d_{1}\otimes d_{2}$) remains an open problem. In this sense, a glimmer of hope comes from the Ref.\cite{Wang2019}, in which the authors emphasize that the closed form of the local quantum uncertainty can be obtained for quantum states of dimension $d_{1}\otimes d_{2}$, by using the generators of the algebra $SU\left(d_{1}\right)$ in optimization as a special unitary group. The explicit expression of the local quantum uncertainty in this case becomes
\begin{equation}
	{\cal U}_{A}\left(\rho \right)=\frac{2}{d_{1}}-\xi_{\max}\left(\hat{\cal W}\right),
\end{equation}
where $\xi_{\max}\left(\hat{\cal W}\right)$ stands for the maximal eigenvalue of the matrix $\hat{\cal W}$ having order $\left(d_{1}^{2}-1 \right)\left(d_{1}^{2}-1 \right)$ and these elements are given by
\begin{equation}
	{\cal W}_{ij}={\rm Tr}\{\sqrt{\rho}\left( \lambda_{i}\otimes\openone_{d_{2}}\right)\sqrt{\rho}\left(\lambda_{j}\otimes\openone_{d_{2}} \right) \}-G_{ij}P,
\end{equation}
with $\lambda_{i}$ are $su\left(d_{1}\right)$ generators. The row vector $G_{ij}$ and the column vector $P$ are given respectively by
\begin{equation}
	G_{ij}=\left(g_{ij1},...,g_{ijk},...,g_{ij{d_{1}^{2}-1}} \right), \hspace{1cm}{\rm avec}\hspace{1cm}g_{ijk}=\frac{1}{4}{\rm Tr}\left(\lambda_{i}\lambda_{j}\lambda_{k}+\lambda_{j}\lambda_{i}\lambda_{k} \right),  
\end{equation}
and
\begin{equation}
	P=\left({\rm Tr}\left(\rho\lambda_{1}\otimes\openone_{d_{2}} \right),...,{\rm Tr}\left(\rho\lambda_{k}\otimes\openone_{d_{2}} \right),...,{\rm Tr}\left(\rho\lambda_{d_{1}^{2}-1}\otimes\openone_{d_{2}} \right) \right)^{T}.
\end{equation}
The other interesting point to note here is that local quantum uncertainty has a geometric interpretation although it only applies to qubit-qudit quantum states. By employing the square root form of a density operator, the (squared) Hellinger distance \cite{Bengtsson2006} between two states $\rho$ and $\varrho$ takes the following form
\begin{equation}
D_{H}^{2}\left(\rho,\varrho\right)=\frac{1}{2}{\rm Tr}\left[ \rho^{1/2}-\varrho^{1/2}\right]^{2}=1-{\rm Tr}\left[\rho^{1/2}\varrho^{1/2} \right].   
\end{equation}
Knowing that $K_{A}=s.\sigma$ is a root-of-unity unitary, then $K_{A}g\left(\rho_{AB}\right)K_{A}=g\left(K_{A}\rho_{AB}K_{A} \right)$ for any bipartite state and for any function $g$. Hence, we can write the skew information in the following fashion
\begin{equation}
{\cal I}\left( {\rho_{AB} ,K_{A}}\right)=1-{\rm Tr}\left[\rho_{AB}^{1/2}K_{A}\rho_{AB}^{1/2}K_{A}\right]=1- {\rm Tr}\left[\rho_{AB}^{1/2}\left(K_{A}\rho_{AB}K_{A} \right)^{1/2} \right]=D_{H}^{2}\left(\rho_{AB},K_{A}\rho_{AB}K_{A}\right). 
\end{equation}
This equation establishes a direct connection between the skew information and the Hellinger distance, indicating that the local quantum uncertainty can be interpreted as the minimum distance between the quantum state before and after a local root-of-unity unitary operation is performed.

\subsection{Interplay between local quantum Fisher information and local quantum uncertainty}
In fact, the notion of local quantum uncertainty (\ref{DLQU}) is very similar to the notion of local quantum Fisher information (\ref{LQFI}). They are two measures based on the concept of quantum uncertainty and both quantify the quantum correlations existing in quantum systems. So it is interesting to study the relationship and the interplay between them. Before providing this relation here, it is first necessary to establish the relation between the skew information (\ref{IS}) and the quantum Fisher information (\ref{QF}) for any qubit-qudit quantum systems. Starting from the above, all mixed quantum states can be written as $\rho=\sum_{n}p_{n}\left|\psi_{n}\right\rangle\left\langle\psi_{n}\right|$, where $\lambda_{n}$ are the eigenvalues of the density matrix $\rho$, and the eigenvectors $\{\left|\psi_{n}\right\rangle\}$ constitute an orthonormal basis such that $\sum_{n}\left|\psi_{n}\right\rangle\left\langle\psi_{n}\right|=\openone$. We can easily show that
\begin{equation}
	{\rm Tr}\left(\rho K^{2} \right)=\sum_{n,m}p_{n}\mid\left\langle\psi_{n} \right|K\left|\psi_{m} \right\rangle\mid^{2},
\end{equation}
which can also be written in the symmetrical form as follows
\begin{equation}
	{\rm Tr}\left(\rho K^{2} \right)=\sum_{n,m}\frac{p_{n}+p_{m}}{2}\mid\left\langle\psi_{n} \right|K\left|\psi_{m} \right\rangle\mid^{2}.
\end{equation}
Then the second term of the equation (\ref{SII}) becomes
\begin{equation}
	{\rm Tr}\left(K\sqrt{\rho}K\sqrt{\rho}\right)=\sum_{n,m}\sqrt{p_{n}p_{m}}\mid\left\langle\psi_{n} \right|K\left|\psi_{m} \right\rangle\mid^{2},
\end{equation}
and the explicit expression of the skew information is given by
\begin{equation}
	{\cal I}\left(\rho,K \right)=\sum_{n,m}\frac{1}{2}\left(\sqrt{p_{n}}-\sqrt{p_{m}} \right)^{2}\mid\left\langle\psi_{n} \right|K\left|\psi_{m} \right\rangle\mid^{2}.
\end{equation}
Here, we mention a non-exhaustive list of properties of the Wigner-Yanase skew information ${\cal I}\left({\rho,K} \right)$, which will be useful in this review \cite{Wigner1963,Luo2003}:
\begin{itemize}
	\item The skew information is always positive, ${\cal I}\left( {\rho ,K} \right)\geq0$, and reduces to the variance if the state $\rho$ is pure ($\rho^{2}\equiv\rho$);
	\begin{equation}
		{\cal I}\left(\left|\psi \right\rangle \left\langle \psi\right|, K \right)={\rm Var}\left(\left|\psi \right\rangle\left\langle\psi \right|,K\right)=\left\langle\psi \right| K^{2}\left| \psi\right\rangle -\left\langle\psi \right| K\left| \psi\right\rangle^{2}.
	\end{equation}
	\item The skew information is always smaller than the variance of the observable $K$. Then 
	\begin{align}
		{\cal I}\left( {\rho ,K} \right)&={\rm Tr}\left(\rho K^{2}\right)-{\rm Tr}\left(\sqrt{\rho}K\sqrt{\rho}K\right)\notag\\&\leq {\rm Tr}\left(\rho K^{2}\right)-{\rm Tr}\left(\rho K\right)^{2}\notag\\&\equiv {\rm Var}\left( {\rho ,K} \right).
	\end{align}
	\item ${\cal I}\left({\rho ,K}\right)$ is a convex function since the skew information decreases when several states are mixed. Then one obtains 
	\begin{equation}
		{\cal I}\left(\sum_{i}p_{i}\rho_{i},K \right)\leqslant\sum_{i} p_{i}{\cal I}\left( \rho_{i},K\right),
	\end{equation}
	for any quantum state $\rho_{i}$, and the constants $p_{i}$ satisfying $\sum_{i}p_{i}=1$ and $0\leqslant p_{i}\leqslant1$. Otherwise, the variance ${\rm Var}\left({\rho,K}\right)$ is a concave function in the state $\rho$.
	\item The skew information is invariant under a unitary transformation $U$, if and only if $U$ and the measured observable $K$ commute, i.e.; $\left[U,K\right]=0$. In this case, one has
	\begin{equation}
		UKU^{-1}=K,\hspace{2cm}\sqrt{U\rho U^{-1}}=U\sqrt{\rho}U^{-1},
	\end{equation}
	and consequently
	\begin{align}
		{\cal I}\left(U\rho U^{-1},K\right)&=-\frac{1}{2}{\rm Tr}\{\left[\sqrt{U\rho U^{-1}},K \right]^{2} \}\notag\\&= -\frac{1}{2}{\rm Tr}\{\left[\sqrt{U\rho U^{-1}},UKU^{-1}\right]^{2} \}\notag\\&=-\frac{1}{2}{\rm Tr}\{U\left[\sqrt{\rho},K\right]^{2}U^{-1}\}\notag\\&=-\frac{1}{2}{\rm Tr}\{\left[\rho,K \right]^{2}\}={\cal I}\left( {\rho ,K} \right).
	\end{align}
	\item Consider a composite system $AB$ defined in the Hilbert space ${\cal H}_{A}\otimes{\cal H}_{B}$, the skew information of the quantum state $\rho_{AB}$ and the reduced density matrix $\rho_{A}$ associated with the subsystem $A$ satisfies the following relation:
	\begin{equation}
		{\cal I}\left(\rho_{AB}, K_{A}\otimes\openone_{B} \right)\geq {\cal I}\left(\rho_{A}, K_{A}\right), 
	\end{equation}
	for all local observables $K_{A}$ acting on the Hilbert space ${\cal H}_{A}$. Here $\openone_{B}$ represents the identity operator in ${\cal H}_{B}$.
	\item In order to obtain an intrinsic quantity providing the information content of the density operator $\rho$, Luo introduced the following average \cite{Luo2006}:
	\begin{equation}
		{\cal Q}_{{\cal I}}\left(\rho \right)=\sum_{i=0}^{n^{2}-1} {\cal I}\left(\rho,K^{i} \right), 
	\end{equation}
	with $\{K^{i}\}$ is a set of $n^{2}$ observables on an $n$-dimensional quantum system. Hence, the global information content of the bipartite state $\rho_{AB}$ via local observables of an $n$-dimensional quantum system $A$ reads
	\begin{equation}
		{\cal Q_{{\cal I}}}_{A}\left(\rho_{AB} \right)=\sum_{i=0}^{n^{2}-1} {\cal I}\left(\rho_{AB},K_{A}^{i}\otimes\openone_{B}\right), 
	\end{equation}
	\item Interestingly, Luo and his co-workers \cite{Luo2012} have shown that the difference between the information content of the state $\rho_{AB}$ and the state $\rho_{A}$, with respect to the local observables $K_{A}^{i}$, can be interpreted as a measure of quantum correlations in the quantum state $\rho_{AB}$. Therefore, we can define an alternative measure of quantum correlations in terms of skew information as
	\begin{equation}
		{\cal F}_{A}\left(\rho_{AB} \right)={\cal Q_{I}}_{A}\left(\rho_{AB} \right)-{\cal Q_{I}}_{A}\left(\rho_{A}\otimes\rho_{B} \right)
	\end{equation}
\end{itemize}
Now, we relate the skew information (\ref{IS}) to the quantum Fisher information (\ref{QF}). Firstly, it is clear that the quantum state $\rho_{\theta}$ satisfies the von Neumann-Landau equation
\begin{equation}
i\frac{\partial\rho_{\theta}}{\partial\theta}=\left[K,\rho_{\theta} \right]. 
\end{equation}
Using the Lyapunov equation (\ref{DLS}), we then find
\begin{equation}
\frac{1}{2}\left(L_{\theta}\rho_{\theta}+\rho_{\theta}L_{\theta}\right)=i\left(\rho_{\theta}K-K\rho_{\theta}\right).  
\end{equation}
Under unitary evolution, the symmetric logarithmic derivative $L_{\theta}$ is independent of the estimated parameter $\theta$ when $K$ and $\rho$ are fixed, i.e., $L=\exp\left(i\theta K\right)L_{\theta}\exp\left(-i\theta K\right)$, implying that
\begin{equation}
	\frac{1}{2}\left(L\rho+\rho L\right)=i\left(\rho K-K\rho\right).  
\end{equation}
Thus, from equation (\ref{QF}) we get
\begin{equation}
{\cal F}_{Q}\left(\rho_{\theta}\right)={\rm Tr}\left(\rho_{\theta}L_{\theta}^{2}\right)={\rm Tr}\left(\exp\left(-i\theta K\right)\rho\exp\left(i\theta K\right)L_{\theta}^{2}\right)={\rm Tr}\left(\rho\exp\left(i\theta K\right)L^{2}\exp\left(-i\theta K\right)\right)={\rm Tr}\left(\rho L^{2}\right).
\end{equation}
Besides, in the spectral decomposition of $\rho$ we have
\begin{equation}
\frac{1}{2}\left\langle \psi_{m} \right|L\rho+\rho L\left| \psi_{n} \right\rangle=\left\langle \psi_{m} \right|i\left[\rho,K \right]\left| \psi_{n} \right\rangle,
\end{equation}
and it is easy to verify that
\begin{equation}
\frac{1}{2}\left(p_{m}+p_{n} \right)\left\langle \psi_{m}\right| L\left| \psi_{n}\right\rangle=i\left(p_{m}-p_{n} \right)\left\langle \psi_{m}\right|K\left|\psi_{n}\right\rangle. 
\end{equation}
In this picture, the quantum Fisher information (\ref{QF}) is given by
\begin{align}
{\cal F}_{Q}\left(\rho,K\right)=\frac{1}{4}\sum_{m,n}p_{m}\mid\left\langle\psi_{m}\right|L \left|\psi_{n}\right\rangle\mid^{2}=\frac{1}{4}\sum_{m,n}p_{n}\mid\left\langle\psi_{m}\right|L \left|\psi_{n}\right\rangle\mid^{2},
\end{align}
which can be written in terms of the eigenvalues and eigenvectors of the density matrix $\rho$ as
\begin{equation}
{\cal F}_{Q}\left(\rho,K\right)=\frac{1}{2}\sum_{m,n}\left(\frac{2\sqrt{p_{m}p_{n}}}{p_{m}+p_{n}}+1\right)\left(\sqrt{p_{m}}-\sqrt{p_{n}}\right)\mid\left\langle\psi_{m}\right|K \left|\psi_{n}\right\rangle\mid^{2}. 
\end{equation}
Since we have
\begin{equation}
0\leq\frac{2\sqrt{p_{m}p_{n}}}{p_{m}+p_{n}}\leq1,
\end{equation}
then the quantum Fisher information is majorized by the Wigner-Yanase skew information and we obtain the above inequalities
\begin{equation}
	{\cal I}\left(\rho,K \right)\leq {\cal F}_{Q}\left(\rho,K\right)\leq2{\cal I}\left(\rho,K \right). \label{183}
\end{equation}
By minimizing the above inequalities (\ref{183}) under a local observable $K_{A}$, we show that the local Fisher quantum information is bounded by the local quantum uncertainty, i.e.,
\begin{equation}
	{\cal U}\left( \rho\right)\leq {\cal Q}_{F}\left(\rho\right) \leq2{\cal U}\left( \rho\right).
\end{equation}
This implies also
\begin{equation}
	{\cal U}\left( \rho\right)\leq {\cal I}\left(\rho,K \right)\leq {\cal F}_{Q}\left(\rho,K \right).
\end{equation}
Therefore, according to the Cramér-Rao theorem (\ref{BCR}), the precision of the parameter $\theta$ can be limited by the local quantum uncertainty (\ref{DLQU}) and by the local quantum Fisher information (\ref{LQFI}) as follows
\begin{equation}
{\rm Var}\left(\rho,\theta \right)=\frac{1}{{\cal F}_{Q}\left(\rho,K\right)}\leq \frac{1}{{\cal Q}_{F}\left(\rho\right)}\leq \frac{1}{{\cal U}\left( \rho\right)}.
\end{equation}
Therefore, the non-classical correlations (local quantum uncertainty and local quantum Fisher information) upper bound the smallest possible variance of the estimator for interferometric phase estimations and thus yield these two inequalities \cite{SlaouiB2019}:
\begin{equation}
	{\rm Var}\left(\theta \right)_{\rm min}\leq \frac{1}{{\cal U}\left( \rho\right)},\hspace{1cm}{\rm and}\hspace{1cm} {\rm Var}\left(\theta \right)_{\rm min}\leq \frac{1}{{\cal Q}_{F}\left(\rho\right)}.
\end{equation}
As result, we conclude here that local quantum uncertainty and local quantum Fisher information both provide tools to understanding the role of non-classical correlations beyond entanglement in improving the accuracy and efficiency in interferometric phase estimations protocols. Moreover, the amount of correlations present in the probe state (the input state) guarantees an upper bound on the smallest possible variance and the minimum achievable statistical uncertainty in the estimated parameter, as quantified by the quantum Fisher information.

\section{Conclusion and comments}			 			
In summary, recent progress in quantum estimation is pushed by the prospect of a second quantum revolution, in order to exploit quantum resources to improve the performance of several quantum information protocols. Here, we have reviewed the main ideas of quantum metrology, with a focus on the role of quantum correlations in quantum metrology, in order to provide a better precision in the estimation of an unknown parameter. Indeed, quantum estimation theory permit the access to the optimal settings that efficiently estimate the value of a parameter. The ultimate limit of the precision of the variance associated with an estimator is bounded according to the quantum Cramér-Rao theorem and is proportional to the inverse of the quantum Fisher information in the case of the estimation of a single parameter and to the inverse of the quantum Fisher matrix in the case of the estimation of several parameters. The basics of classical and quantum estimation theory are discussed. In addition, analytical methods for computing these two quantities in several scenarios are discussed.\par

We also discuss recent advances concerning the interpretation of non-classical correlations in quantum metrology, by using the quantifiers that are based mainly on the uncertainty of local observables over the quantum states. Whenever a quantum system shares quantum correlations, quantified by local quantum uncertainty and local quantum Fisher information, quantum mechanics predicts that any local measurement has a degree of uncertainty that gives us an improved sensitivity in an interferometric phase estimation scheme. These two discord-type quantifiers quantify the minimum amount of accuracy of the estimated parameters. An interesting question here is to investigate whether these metrological measures of quantum discord can be extended to multipartite systems to improve the performance of protocols in quantum metrology. Furthermore, can we find the interaction between quantum coherence \cite{Streltsov2017} and quantum estimation theory and analyze the effects of decoherence on the sensitivity in quantum metrology? On the other hand, many quantum multiparametric estimation schemes have been proposed and discussed for various quantum systems, and some of them have shown theoretical advances over single parameter schemes. However, many problems remain open, such as the conception of an optimal measure, a particularly simple and practical measure that is independent of the unknown parameters, analytical methods to saturate the Cramér-Rao bound as well as the role of quantum correlations in the quantum multiparametric estimation strategy. Progress in these subjects will be reported elsewhere.

\end{document}